\documentclass[conference]{IEEEtran}
\IEEEoverridecommandlockouts

\usepackage[frozencache,cachedir=.]{minted}

\usepackage{tablefootnote}
\usepackage{algorithmic}
\usepackage{graphicx}
\usepackage{textcomp}
\usepackage[table]{xcolor}
\usepackage{minted}
\usepackage{hyperref}
\usepackage{multirow}
\usepackage{enumitem}           % for numbered enumerations and margins
\usepackage[T1]{fontenc} % for extra fonts 
\usepackage{tikz} % for paper drawings
\usepackage{stfloats}
\usepackage[utf8]{inputenc}
\usepackage{soulutf8}
\usepackage{booktabs}
\usepackage{colortbl}
\definecolor{mygray}{rgb}{0.71,0.71,0.71}
\usepackage{balance}
\usepackage{amsfonts}
\usepackage[caption=false]{subfig} % https://tex.stackexchange.com/questions/169475/referencing-in-subfloat-not-working/169477#169477
\usepackage{array}
\usepackage{cite}
\newcolumntype{P}[1]{>{\centering\arraybackslash}p{#1}}
\newcommand{\shortref}[1]{$\S$~\ref{#1}}

\newcommand{\issue}[1]{\textsc{Q.I.{#1}}}

\newcommand{\ie}{\textit{i.e.},\ }
\newcommand{\eg}{\textit{e.g.},\ }
\newcommand{\etal}{\textit{et al.} }
\newcommand{\etc}{{\em etc.}}
\newcommand{\concept}[1]{\textbf{\small\textsf{#1}\normalsize}}

\newcommand{\green}[1]{\textcolor{teal}{\textbf{#1}}}

\usepackage{booktabs}

\definecolor{francBlue}{RGB}{64,76,87}

\usepackage[most]{tcolorbox}
\usepackage[parfill]{parskip}

\tcbset{
  parskip/.style={
    before={\par\pagebreak[0]\parindent=0pt},
    after={\parfillskip=0pt\par},
  },
}
\newtcolorbox{resultbox}[1][]{%
    colback=black!3,
    colframe=black!3,
    notitle,
    sharp corners,
    borderline west={2pt}{0pt}{gray!80!black},
    enhanced,
    breakable,
    boxsep=0pt,
    left=4pt,right=2pt,top=2pt,bottom=2pt,
    }

\definecolor{codebg}{rgb}{0.99,0.99,0.99}
\definecolor{hiliteColor}{rgb}{1,0.92549019607,0.6}
\definecolor{tainted}{rgb}{0,1,1}

\newmintinline{python}{fontsize=\normalsize}

\newminted[PythonSourceCode]{python}{
    labelposition=topline,
    frame=single,
    numbersep=2pt,
    fontsize=\tiny,
    xleftmargin=5pt,
    framerule=0.5pt,
    linenos
}

\newminted[JavaSourceCode]{java}{
    labelposition=topline,
    frame=single,
    numbersep=2pt,
    fontsize=\tiny,
    xleftmargin=5pt,
    framerule=0.5pt,
    linenos
}

\definecolor{magnolia}{rgb}{0.97, 0.96, 1.0}
\definecolor{shadecolor}{rgb}{0.97, 0.96, 1.0}
\newcommand{\code}[1]{\texttt{\small{#1}\normalsize}}
\newmintinline[codePython]{python}{fontsize=\small}
\newmintinline[codeJava]{java}{fontsize=\small}
\newmintinline[snippetJava]{java}{fontsize=\small,bgcolor=magnolia}
\newmintinline[snippetPython]{python}{fontsize=\small,bgcolor=magnolia}
\newmintinline[snippetPerl]{perl}{fontsize=\small,bgcolor=magnolia}

\newcommand{\joanna}[1]{\textcolor{magenta}{}}

\newcommand{\subparagraph}[1]{\noindent\textbf{#1}\quad}

\AtBeginDocument{%
  }

\begin{document}

% \title{The Fault in our Stars: Quality Assessment of Prompts Used in Code Generation Benchmarks}
% \title{Quality Assessment of Prompts Used in Code Generation}
% \title{The Fault in our Stars: Quality Assessment of Datasets for Code Generation Evaluation}

\title{The Fault in our Stars: Quality Assessment of Code Generation Benchmarks}

\author{%
\IEEEauthorblockN{Mohammed Latif Siddiq}
\IEEEauthorblockA{\textit{University of Notre Dame} \\
Notre Dame, IN. USA \\
msiddiq3@nd.edu}
\and
\IEEEauthorblockN{Simantika Dristi\textsuperscript{\textsection}, Joy Saha\textsuperscript{\textsection}}
\IEEEauthorblockA{\textit{University of Virginia} \\
Charlottesville, VA. USA \\
\{nwc8gr, daa7mv\}@virginia.edu}
\and
\IEEEauthorblockN{Joanna C. S. Santos}
\IEEEauthorblockA{\textit{University of Notre Dame} \\
Notre Dame, IN. USA \\
joannacss@nd.edu}
}
% \author{Simantika Dristi and Joy Saha}
% \email{{simantika.dristi,joy.saha}@bracu.ac.bd}
% \affiliation{%
%   \institution{BRAC University}  
%     \city{}
%   \state{}
%   \country{}
%   \postcode{}
% }
% \author{Joanna C. S. Santos}
% \email{joannacss@nd.edu}
% \affiliation{%
%   \institution{University of Notre Dame} 
%   \city{Notre Dame}
%   \state{IN}
%   \postcode{46556}
% }

% \author{
%     \IEEEauthorblockN{Mohammed Latif Siddiq\IEEEauthorrefmark{1}, Simantika Dristi\IEEEauthorrefmark{2}\textsuperscript{\textsection},  Joy Saha\IEEEauthorrefmark{2}\textsuperscript{\textsection}, Joanna C. S. Santos\IEEEauthorrefmark{1}}
%     \IEEEauthorblockA{\IEEEauthorrefmark{1}Department of Computer Science and Engineering, University of Notre Dame, USA}
%     \IEEEauthorblockA{\IEEEauthorrefmark{2}Department of Computer Science, University of Virginia, USA
%     \\msiddiq3@nd.edu, \{nwc8gr, daa7mv\}@virginia.edu, joannacss@nd.edu}
% }

% \renewcommand{\shortauthors}{Siddiq et al.}


\maketitle

 \begingroup\renewcommand\thefootnote{\textsection}
 \footnotetext{These authors equally contributed to this work.}
 \endgroup

\begin{abstract}
Large Language Models (LLMs) are gaining popularity among software engineers.
A crucial aspect of developing effective code generation LLMs is to evaluate these models using a robust benchmark. Evaluation benchmarks with quality issues can provide a false sense of performance. In this work, we conduct the first-of-its-kind study of the quality of prompts within benchmarks used to compare the performance of different code generation models. To conduct this study,  we analyzed 3,566 prompts from 9 code generation benchmarks to identify quality issues in them. We also investigated whether fixing the identified quality issues in the benchmarks' prompts affects a model's performance. We also studied memorization issues of the evaluation dataset, which can put into question a benchmark's trustworthiness. We found that code generation evaluation benchmarks mainly focused on Python and coding exercises and had very limited contextual dependencies to challenge the model. These datasets and the developers' prompts suffer from quality issues like spelling and grammatical errors, unclear sentences to express developers' intent, and not using proper documentation style. Fixing all these issues in the benchmarks can lead to a better performance for Python code generation, but not a significant improvement was observed for Java code generation. We also found evidence that GPT-3.5-Turbo and CodeGen-2.5 models may have data contamination issues.
\end{abstract}

%    <concept>
%        <concept_id>10011007.10010940.10011003.10011002</concept_id>
%        <concept_desc>Software and its engineering~Software performance</concept_desc>
%        <concept_significance>500</concept_significance>
%        </concept>
%    <concept>
%        <concept_id>10011007.10010940.10011003.10011687</concept_id>
%        <concept_desc>Software and its engineering~Software usability</concept_desc>
%        <concept_significance>500</concept_significance>
%        </concept>
%    <concept>
%        <concept_id>10011007.10011074.10011099.10011693</concept_id>
%        <concept_desc>Software and its engineering~Empirical software validation</concept_desc>
%        <concept_significance>300</concept_significance>
%        </concept>
%  </ccs2012>

% \ccsdesc[500]{Software and its engineering~Software performance}
% \ccsdesc[500]{Software and its engineering~Software usability}
% \ccsdesc[300]{Software and its engineering~Empirical software validation}


\section{Introduction}

Code generation models generate code by taking as input a \textit{prompt}, which captures the developers' intent~\cite{Le_2021}. These models are increasingly popular among software developers~\cite{perry2022users}. In fact, a recent survey with 500 US-based developers who work for large-sized companies showed that \textbf{92\%} of them are using AI-based code generation tools both for work and personal use~\cite{shani2023survey}. Part of this fast widespread adoption is due to the increased productivity perceived by developers~\cite{research_copilot}; AI helps them to automate repetitive tasks so that they can focus on higher-level challenging tasks~\cite{albert22productivity}. 
% \textbf{Not only are developers using AI code assistants, but so are our students}~\cite{perry2022users,ernst2022aide}. 

As code generation models are becoming ubiquitous during software development~\cite{ernst2022aide}, 
% it is vital that the generated code is free of vulnerabilities for two main reasons. First, insecure generated code can get deployed into production, leading to vulnerabilities. Second, these tools would be teaching students insecure coding practices. Both are undesirable outcomes. 
% As machine-learning-based automated code generation is advancing at an unprecedentedly fast pace, 
the need for reliable \textit{\textbf{evaluation benchmarks}} is vital. Code generation benchmarks are crucial for evaluating and comparing the effectiveness of various models in producing code. These benchmarks are designed to assess the generated codes from various perspectives, such as their \textit{correctness}, \textit{readability}, and \textit{security}~\cite{zan2023NL2Code}.

% More than 30 evaluation datasets exist for code generation~\cite{zan2023NL2Code}. Benchmarking the code generation models heavily depends on the evaluation dataset, but they may not have good quality.
% Code generation tools like GitHub Copilot \cite{copilot} can improve productivity according to their internal research \cite{research_copilot}, and it also shows the developers can be 55\% faster than the developers who do not use GitHub Copilot. The usage of and trust in these code-generation model-based tools depends on the backend models' performance \cite{cheng2022would}, and they are evaluated with various benchmarks.

While there are over \textbf{15} evaluation benchmarks for code generation models~\cite{zan2023NL2Code}, their \textit{quality} and \textit{reliability} are currently unclear. First, these benchmarks are often collected in an \textit{ad-hoc} fashion, which may not be representative of real software scenarios~\cite{CoderEval}. Second, as these benchmarks are curated from publicly available data, there is the risk that existing models include data from these benchmarks in the training set (\ie \textit{test set contamination}~\cite{carlini2019secret}). In this case, the reliability of \texttt{pass@k}~\cite{HumanEval} and other performance metrics is put into question, as the models might have memorized the solutions to the prompts in the dataset~\cite{sallou2023breaking}. 
Therefore, issues on these benchmarks significantly impact the trustworthiness of the evaluation results, making it crucial to thoroughly investigate and evaluate the benchmarks themselves. 

% developers' prompts and code generation benchmarks. 
In light of this research gap, 
% this empirical study aims to delve into the intricacies of code generation benchmarks, seeking to provide insights into their accurate measure of quality.
we present an \textbf{\textit{empirical study of the quality of prompts in benchmarks from multiple dimensions and compare them with issues observed in real world prompts created by developers}}. Specifically, we systematically analyzed \textbf{3,566} prompts from \textbf{9} Java and Python benchmarks to check the quality issues in these prompts. We observed that issues fall into three categories: formatting, a prompt containing sentences that do not properly (or incorrectly) specify the behavior of the code, and  prompts containing unnecessary tokens (\textit{noise}).
Besides identify quality issues in the benchmarks' prompts, we also explored to what extent these issues affect a model's performance. In doing so, we found that fixing spelling and grammatical issues and using standard JavaDoc and docstring style can help models to generate code.
Last, but not least, we studied whether existing models are memorizing answers from existing benchmarks (\ie test set contamination~\cite{sallou2023breaking,carlini2023quantifying}). In our experiments, we found empirical evidence of testset contamination in two models: \textsc{CodeGen-2.5} and \textsc{GPT-3.5}.

The contributions of this paper are:
\begin{itemize}[leftmargin=*,label=--,noitemsep,topsep=0pt]
    \item  A thorough investigation of code generation benchmarks' prompts (\textbf{RQ1} and \textbf{RQ2}) so researchers and developers can make informed decisions about choosing a benchmark to evaluate code generation models.
    \item A study of how fixing quality issues in a prompt can affect a model's evaluation (\textbf{RQ3}).
    \item An investigation of possible test-set contamination issues in HumanEval, a popular benchmark (\textbf{RQ4}). 
    \item A comparison of the quality issues observed in benchmarks' prompts and the  real world prompts made by developers when interacting with ChatGPT \cite{chatgpt} (\textbf{RQ5}).

    % This investigation will help researchers identify gaps and shortcomings in existing benchmarks, paving the way for improved and more accurate evaluation frameworks.

    % \item 

\end{itemize}
This paper's {replication package} is  available in \url{https://github.com/s2e-lab/Datasets-Quality}.

\section{Background}

% This section presents key concepts to understand the paper.
% This section defines key concepts to understand the paper.
\subsection{Large Language Models}\label{subsec:LLM}

\concept{{Large Language Models}} (\concept{LLMs})~\cite{LLM}  are neural networks with tens of millions to billions of parameters that were trained on large amounts of unlabeled text using self-supervised or semi-supervised learning \cite{brown2020language}.
LLMs are intended to be general purpose models for many natural language processing tasks, such as text generation, translation, summarization, \etc~ 
% \textsf{BERT} (\textit{Bidirectional Encoder Representations from Transformers}) \cite{bert2018}, \textsf{T5} (\textit{Text-to-Text Transformer}) \cite{2020t5} and \textsf{GPT-3} (\textit{Generative Pre-trained Transformer}) \cite{brown2020language} are examples of LLMs.
While LLMs are trained to understand \textit{natural} language, they can be fine-tuned with source code samples to understand \textit{programming} languages.  This makes LLMs useful for a myriad of software engineering tasks, such as code completion \cite{izadi2022codefill,kim2021code,svyatkovskiy2021fast,codebert}, and summarization \cite{gao2022m2ts}. {CodeBERT} \cite{codebert}, {CodeT5} \cite{codet5}, and {Codex} \cite{HumanEval} are examples of ``\concept{code LLMs}'', \ie LLMs trained on source code (henceforth, simply ``{LLMs}'').  
% \concept{Code LLMs} (henceforth, simply ``{LLMs}'') generate code from a given 

Given a \textit{prompt} as input, a code LLM  generates code tokens, one by one, until it reaches a \textit{stop sequence} (\ie a pre-configured token sequence) or the \textit{maximum number of tokens} is reached. A \concept{prompt}  provides a high-level specification of a developer's intent and can include different code elements, \eg function signatures, expressions, comments, \etc 

% , as in the case of BERT \cite{bert2018,codebert},
Transformer-based code generation models employ masked language modeling objectives or \textit{left-to-right} (causal) autoregressive language modeling objectives \cite{brown2020language,HumanEval}. That is, to generate  code, the generative model will use the context on the \textit{left} side of the cursor and ignore any context on the \textit{right}. 
% Nevertheless, developers alter code in earlier sections, and such alteration might rely on both sides of the cursor, not only the left side. 
The process of creating code that incorporates context from \textit{both} sides is known as \concept{code infilling}. In this work, we focus on studying \concept{left-to-right} code generation benchmarks because the majority of benchmarks are meant to evaluate left-to-right code generation~\cite{zan2023NL2Code}. For instant, we considered 9 benchmarks in our study out of 17 benchmark studied in this survey by Daoguang \etal \cite{zan2023NL2Code}. 
% \canRemove{Code infilling mimics developers' edit actions in an existing code, which is an action that may depend on both sides of the cursor.}
% One illustration of a code-infilling model is InCoder \cite{incoder}, and StarCoder \cite{StarCoder}.

% Transformer-based code generation techniques use left-to-right (causal) autoregressive language modeling objectives \cite{brown2020language,HumanEval} or, as BERT does \cite{bert2018,codebert}, use masked language modeling objectives. That means that the generative model will take context from the \textit{left} side of the cursor and will not take any context \textit{after} the cursor to generate source code. However, developers edit earlier parts of the code, and this edit action may depend on both sides of the cursor. Generating code by taking context from \textit{both} sides is referred to as \textit{\textbf{code infilling}}. 
 % It is a code generation technique that takes context from both sides and not only from the left side. 
% InCoder \cite{incoder} is an example of a code-infilling model.

\subsection{Code Generation Benchmarks}
\concept{Code generation benchmarks} are used to evaluate and compare models based on different metrics~\cite{HumanEval, MBXP}. Existing benchmarks usually contain coding problems captured in a natural language, comment, or combination of comment and code, referred to as a \concept{prompt} \cite{Le_2021}. After using the prompt for code generation, different metrics can be used to evaluate the performance. For example, \texttt{CodeBleu} \cite{ren2020codebleu} can be used for syntactical correctness, and \texttt{pass@k} \cite{HumanEval} can be used for functional correctness. Benchmarks may be created for a specific purpose. For example, SALLM \cite{siddiq2023generate} focuses on evaluating the security of generated code and uses the \texttt{vulnerable@k} to compare the performance of the models.
% The generated codes can be evaluated in different ways, such as by comparing with a ground truth code based on their syntactical and data flow matching \cite{hao2022aixbench} or running the code with pre-defined tests~\cite{HumanEval}. 

\subsection{Memorization in LLMs}

\concept{Memorization} refers to a model's ability to preserve and generate an identical string from its training data~\cite{carlini2019secret,carlini2021extracting}. 
In this work, we used a similar definition as the one presented by Carlini~\etal~\cite{carlini2021extracting}. Specifically, if there exists a prompt that generates a code snippet that completely matches any of its training data code snippets, then this code snippet is considered to be memorized by the code generation model, a case of \concept{verbatim memorization}~\cite{zhou2023quantifying}. 

In light of this definition and similar to prior work~\cite{yang2024unveiling}, we study whether test set contamination by verifying whether the generated code is a \textit{clone} of the solution available in the benchmark. Specifically, in our work, we search for \textit{type-1}, \textit{type-2}, and \textit{type-3} code clones to pinpoint memorization~\cite{ROY2009470}. 
A \concept{type 1} clone occurs when two code snippets have identical code fragments except for layout, comments, and whitespace differences. A \concept{type 2} clone arises when there are syntactically identical fragments except for comments, whitespace, literals, identifiers, and types. A 
\concept{type 3} clone means that there are copied fragments that have undergone additional changes, such as additions, deletions, or changes to statements, as well as adjustments to identifiers, literals, types, whitespace, layout, and comments.

\section{Methodology}\label{sec:Methodology}

In this paper, we answer the following questions:

\begin{itemize}[leftmargin=22pt,label=-,noitemsep,topsep=0pt]
    \item[\small \bf RQ1] \textit{\textbf{How representative are existing benchmarks of real-world code generation usage scenarios?}} 
\end{itemize}
Code generation models need rigorous evaluation and verification. However,  existing benchmarks may not represent real-world scenarios and cover many programming languages. In this question, we compare the code generation benchmarks' \textit{covered programming languages}, \textit{usage scenario(s)}, \textit{number of prompts}, and \textit{contextual dependency complexity}.

\begin{itemize}[leftmargin=22pt,label=-,noitemsep,topsep=0pt]
    \item[\small \bf RQ2] \textit{\textbf{What are the quality issues in the prompts within code generation benchmarks?}} 
\end{itemize}
In this RQ, we study quality issues in the prompts of code generation benchmarks.
To do so, we manually analyzed a total of \textbf{3,566} from \textbf{9} code generation benchmarks. We performed \textit{open coding}~\cite{stol2016grounded} of the prompts in these benchmarks to identify and categorize quality issues.

% Given that Existing benchmarks suffer from various quality issues, such as inconsistent evaluation criteria, insufficient diversity in datasets, and inadequate scalability, which can compromise the reliability and generalizability of their results.

 % across three different dimensions: the prompt's \textit{intent}, \textit{formatting}, and \textit{noise}. We manually analyzed 

\begin{itemize}[leftmargin=22pt,label=-,noitemsep,topsep=0pt]
    \item[\small \bf RQ3] \textit{\textbf{Does improving the quality of a prompt in code generation benchmarks affect the evaluation result?}}
\end{itemize}
    % Quality issues in the datasets' prompts can negatively impact the benchmarking result. 
    
    We investigate whether improving the quality of  a benchmark's prompts  affects the results of code generation models. To do so, we  fixed quality issues identified in \textbf{RQ2} and compared the performance of LLMs when given as input the \textit{fixed} prompts and the \textit{original} prompts with quality issues.

\begin{itemize}[leftmargin=22pt,label=-,noitemsep,topsep=0pt]    
    \item[\small \bf RQ4] \textit{\textbf{Are there contamination issues in existing code generation benchmarks?}}
\end{itemize}
    Since code generation models are fine-tuned on source code from open-source repositories, there is a risk that code from  evaluation benchmarks are in the models' training set. If prompts from benchmarks are in the training set, the code generation model can perform better because it has memorized the answer~\cite{sallou2023breaking}. Hence, this contamination issue will affect the code generation model's benchmarking process. In this RQ, we explore the possibility of contamination issues in  existing code generation benchmarks.

\begin{itemize}[leftmargin=22pt,label=-,noitemsep,topsep=0pt]

\item[\small \bf RQ5] \textit{\textbf{Are the quality issues in the benchmarks' prompts similar to issues observed in real world prompts?}} 
\end{itemize}
In this RQ, we explore whether the quality issues in the benchmarks' prompts are similar to the ones that are observed in the real world, \ie from developers using LLMs in their day-to-day development activities. To answer this question, we extracted prompts from 
% Developers are using ChatGPT in their day-to-day life for learning new frameworks, refactoring, explaining code blocks \etc \cite{devgpt, siddiq2024devgpt}. They asked the ChatGPT \cite{chatgpt} various questions, but the style of asking can be problematic while they are shared with other developers. In this research question, we investigate different quality issues of the prompts crafted by real-world developers.
\textit{DevGPT}~\cite{devgpt}, a dataset that contains the chats from  software developers  interacting with ChatGPT~\cite{chatgpt}. This dataset was curated by finding \textit{ChatGPT share links} that were posted on GitHub \textit{issues}, \textit{pull requests}, \textit{discussions}, \textit{commits}, \textit{code files}, and \textit{threads} on Hacker News.

We detail how we answer each RQ in the next sections.

% We describe how we answer each of these RQs in the next sections.

\subsection{RQ1: Code Generation Benchmarks Comparison}

To answer RQ1, we first  collected 17 benchmarks listed in a recent survey \cite{zan2023NL2Code}. 
% In addition, we collected \hl{X} datasets using HuggingFace search~\cite{HuggingFace}. HuggingFace is a website where researchers and practitioners release their models and dataset publicly. We also snowballed by checking recent papers on code-generation tasks from PaperswithCodes~\cite{PapersWithCode}. We collected \textbf{Y} evaluation datasets from this website. 
Since we focus on \textit{left-to-right} code generation, we disregard benchmarks designed to evaluate code-infilling models. This way, we obtained a total of \textbf{9} benchmarks: 
\textsc{MXEVAL}~\cite{MBXP}, 
\textsc{CoderEval}~\cite{2023arXiv230200288Y},
\textsc{ODEX}~\cite{wang2023executionbased}, 
\textsc{MBPP}~\cite{MBPP},
\textsc{TorchDataEval}~\cite{zan2022language},
\textsc{HumanEval}~\cite{HumanEval},
\textsc{PandasEval}~\cite{zan2022cert},
\textsc{NumpyEval}~\cite{zan2022cert}, and
\textsc{JigsawDataset}~\cite{jain2021jigsaw}.
\textsc{MXEVAL}~\cite{MBXP} is a benchmark that extends the \textsc{MathQA}~\cite{MathQA}, \textsc{MBPP}~\cite{MBPP}, and \textsc{HumanEval}~\cite{HumanEval} benchmarks.

To verify a benchmark's potential of representing real-world code generation scenarios, we analyzed each benchmark  to identify their \textbf{(i)} \textit{\textbf{covered programming language(s)}}, \textbf{(ii)} \textit{\textbf{usage scenarios}}, \textbf{(iii)} \textit{\textbf{number of prompts}}, and \textbf{(iv)} \textit{\textbf{contextual dependency complexity}}.
We classify a benchmark's \textit{contextual dependency complexity} based on the categorization scheme described by Yu~\etal~\cite{CoderEval}. This complexity can be:

\begin{itemize}[leftmargin=*,label=-,noitemsep,topsep=0pt]
    \item \textbf{self-contained}: benchmarks whose solution to the prompt can be implemented using only built-in classes/modules that do not need to be imported (\eg  Java's String class does not need to be imported to be used).
    \item \textbf{slib-runnable}: benchmarks where the solution to the prompts needs to import classes/modules that are provided by the language and do not require further installation (\eg Java's \codePython{java.util} package and Python's \codePython{re} module).
    \item \textbf{plib-runnable}: benchmarks in which the prompts' solutions only use libraries that are publicly available on PyPi or Maven central (\eg Apache Log4j).
    \item \textbf{class-runnable}: benchmarks in which the solution uses code elements (\eg methods, objects) that are declared outside the prompt's method but within the prompt's class.
    \item \textbf{file-runnable}: benchmarks in which the generated solution uses code elements \textit{outside} its class, but that is still declared on the same file as the prompt.
    \item \textbf{project-runnable}: benchmarks in which the generated code uses code elements declared in other source files in the benchmark.
\end{itemize}

To answer RQ1, we created a \textit{\textbf{benchmark profile}} identifying the information above for each benchmark. This benchmark profile was created by examining the  benchmarks' original paper and technical documentation to identify the metadata \textbf{(i)}--\textbf{(iv)} listed above.

\subsection{RQ2: Benchmark Quality Evaluation}\label{subsec:RQ2Methodology}

Since there were over 8,000 prompts in total in the studied benchmarks, we first randomly sample prompts from each chosen benchmark with a \textbf{99\%} confidence interval and a \textbf{5\%} margin of error. As shown in Table~\ref{tab:dataset-samples}, we analyzed a total of \textbf{3,566} prompts from \textbf{9} benchmarks. 

\begin{table}[!ht]
\setlength{\aboverulesep}{0pt}
\setlength{\belowrulesep}{0pt}
\setlength{\extrarowheight}{.75ex}
\setlength\tabcolsep{0.5pt} % default value: 6pt
\centering
\caption{Total number of prompts and sampled prompts per benchmark.}
\label{tab:dataset-samples}
% \footnotesize\vspace{-10pt}
\rowcolors{3}{gray!5}{white}
\begin{tabular}{@{}lcclcc@{}}
\toprule
\rowcolor[HTML]{efefef} 
\multicolumn{1}{c}{\cellcolor[HTML]{efefef}\textbf{Benchmark}} & 
\textbf{\begin{tabular}[c]{@{}c@{}}\#\\Prompts\end{tabular}} & 
\textbf{\begin{tabular}[c]{@{}c@{}}\#Sampled\\Prompts\end{tabular}} &  
\multicolumn{1}{c}{\cellcolor[HTML]{efefef}\textbf{Benchmark}} & 
\textbf{\begin{tabular}[c]{@{}c@{}}\#\\Prompts\end{tabular}} & 
\textbf{\begin{tabular}[c]{@{}c@{}}\#Sampled\\Prompts\end{tabular}}
\\ \midrule
\textbf{MXEVAL}~\cite{MBXP} & 6,031 & 2,037 & \textbf{MBPP}~\cite{MBPP} & 426 & 261 \\
\hspace{9pt}\textbf{\footnotesize{MBPP}} & 1,940 & 791 & \textbf{TorchDataEval}~\cite{zan2022language} & 302 &  270\\
\hspace{9pt}\textbf{\footnotesize{HumanEval}} & 325 & 262 &  \textbf{HumanEval}~\cite{HumanEval} & 164 & 132    \\ 
\hspace{9pt}\textbf{\footnotesize{MathQA}} & 3,766 & 984 & \textbf{PandasEval}~\cite{zan2022cert} & 101  & 88 \\
\textbf{CoderEval}~\cite{2023arXiv230200288Y} & 460 & 342 & \textbf{NumpyEval}~\cite{zan2022cert} & 101 & 88 \\
\textbf{ODEX}~\cite{wang2023executionbased} & 439 & 265  &  \textbf{JigsawDataset}~\cite{jain2021jigsaw} & 88 & 83  \\
\bottomrule
\end{tabular}
\end{table}

% To answer this RQ, we first collected popular code generation benchmarks (\shortref{sec:BenchmarkCollection}) and systematically analyzed their prompts in order to identify \textit{quality issues} (\shortref{sec:QualityIssues}).

% \subsubsection{Quality Issues}\label{sec:QualityIssues}

% For each sampled prompt, we focused on identifying quality issues related to \textit{\textbf{intent}}, \textit{\textbf{formatting}}, and \textit{\textbf{noise}}.

% \checkThis{As the results in RQ2 will show (\shortref{sec:ResultsRQ2}), these benchmarks mostly supported  Python and Java. Thus, 

In our study, we focused on Java and Python prompts because not only these are popular languages among developers \cite{so2023survey}, but they are also the most supported language in benchmark datasets (\shortref{sec:ResultsRQ1}). Thus, we systematically analyzed the benchmarks' Python/Java prompts to identify \textit{quality issues}. This qualitative analysis was performed by two of the authors,  with over two years of software development and teaching experience each. Each author independently performed open coding~\cite{stol2016grounded} of each prompt.
The open coding started with a (initially empty) shared ``code book'' where we progressively captured the issue's \textit{title} and \textit{description} with \textit{examples} as we analyzed prompts. Our code book was constantly refined throughout the open coding process. 

After each author finished the open coding, a third author, who has over three years of professional programming experience, resolved the discrepancies through discussion and mediation. This analysis took us approximately \textbf{650} person-hours. 
We calculated the Cohen’s Kappa coefficient to measure the inter-rater reliability of this analysis, and it was \textbf{0.76}, which indicates \textbf{\textit{substantial agreement}} \cite{mchugh2012interrater}.

% The Cohen’s Kappa coefficient \cite{mchugh2012interrater} that measures our inter-rater reliability was \textbf{0.76}, which indicates \textbf{\textit{substantial agreement}} \cite{mchugh2012interrater}.
% \hl{Bring Cohen's Kappa here}
% with an average of 1 year of professional programming experience

\subsection{RQ3: Impact on Performance}
In RQ2, we identified the quality issues in the benchmarks' prompts. In RQ3, we fixed the issues to verify  to what extent fixing them affects the performance of the models. As it would be time-consuming to manually fix thousands of  prompts, we fixed the issues identified for the Python and Java version of the \textsc{HumanEval} benchmark~\cite{HumanEval,MBXP}. We chose this benchmark because most of the code generation models are evaluated with it, as shown in a popular leaderboard published on PapersWithCode.com~\cite{paperswithcode} which lists over 120 code LLMs that were evaluated with \textsc{HumanEval}.

To conduct this investigation, the same two authors who have done the open coding in RQ2 went through all the issues identified in RQ2 for the \textbf{164} prompts from \textsc{HumanEval} Python \cite{HumanEval} and \textbf{161} prompts from \textsc{HumanEval}'s Java version \cite{MBXP}.  For each identified issue, we created a set of \textit{fix guidelines} that was shared among both researchers. Since a prompt can have \textit{more than one} quality issue, the authors first fixed the issues \textit{one by one} by creating a modified prompt version that does not contain \textit{one} particular type of problem. Subsequently, they created a new prompt version that fixed \textit{all} the quality issues in the prompt. Once both authors fixed all prompts, they peer-reviewed each other's fixes and came up with a final fix. To mitigate subjectivity when fixing issues,  the senior author, who has over 10 years of experience, checked the modification and updated the prompts in multiple rounds of discussions. After that, we have 501 prompts for Java and 663 prompts for Python, including the original prompts.

% As we will detail in \shortref{sec:ResultsRQ2}, we have \textbf{663} and \textbf{501} prompts for the Python and Java version of HumanEval, respectively, including the original prompts with quality issues (\ie \textbf{164} Python and \textbf{161} Java prompts). There are 27 prompts fixed for issue I1, 1 prompt for I4, 7 prompts for ID I5, 1 prompt for I6, 164 prompts for I9, 147 prompts for I10, and 164 prompts after fixing all problems for the HumanEval Python. For Java, there are 22 prompts fixed for issue I1, 6 prompts for I5,1 prompt for each of ID P6 and P7, 161 prompts for I9, 67 prompts for I10, and 161 prompts after fixing all problems. 

After fixing these issues, we give the prompts (\textit{fixed} and \textit{original} ones) to five code generation LLMs:

% We used five LLMs to generate code from these prompts:

\begin{itemize}[leftmargin=*,label=-,noitemsep,topsep=0pt]
    \item \textbf{\textsc{CodeGen}} ~\cite{Nijkamp2022ACP} is a code LLM that has three variants: \textit{CodeGen-nl}, \textit{CodeGen-multi}, and \textit{CodeGen-mono}. \textsc{CodeGen-nl}, trained with the \textit{Pile} dataset~\cite{ThePileDataset}, is focused on text generation. The \textit{CodeGen-multi} is built on top of \textit{CodeGen-nl} but further trained with a large scale-dataset of code snippets in six different languages (\ie C, C++, Go, Java, JavaScript, and Python)~\cite{BigQueryDataset}. The \textit{CodeGen-mono} is built from \textit{CodeGen-multi} and further trained with a dataset of only Python code snippets~\cite{Nijkamp2022ACP}. They also released another (newer) version called \textit{CodeGen-2.5} \cite{nijkamp2023codegen2}, which is trained on the StarCoder data from BigCode \cite{Kocetkov2022TheStack}. It has a mono and multi-version. We use \textbf{\textsc{CodeGen-2.5-7B-mono}} to generate Python code and \textbf{\textsc{CodeGen-2.5-7B-multi}} to generate Java code.

    \item \textbf{\textsc{SantaCoder}} \cite{allal2023santacoder} is a 1.1B parameter LLM trained on the Java, JavaScript, and Python subsets of The Stack \cite{Kocetkov2022TheStack} dataset. It can do both \textit{left-to-right} generation and \textit{infilling}. 

    \item \textbf{\textsc{StarCoder}}~\cite{StarCoder}  is an LLM with 15.5B parameters trained with over 80 different programming languages. This model is focused on fill-in-the-middle objectives and can complete code given a code-based prompt. It has two versions for code generation: \textit{StarCoderBase} and \textit{StarCoder}. As the latter one is further trained with Python samples, we used that for Python and the former for Java code generation. 

    \item \textbf{\textsc{WizardCoder}} \cite{luo2023wizardcoder} is an instruct-tuned version of \textsc{StarCoder}~\cite{StarCoder} model using Evol-Instruct method on the code domain. This model can generate both code and follow complex instructions.
    
    \item The \textbf{\textsc{Generative Pre-trained Model (GPT)}} ~\cite{brown2020language} is a family of  transformer-based~\cite{attention2017} and task-agnostic LLMs that can \textit{understand} and \textit{generate} natural language. We used \textbf{\textsc{GPT-3.5-Turbo}},  which is tuned for chat-style conversation and powers a popular chat-based question-answering tool, ChatGPT \cite{chatgpt}.
\end{itemize}

We chose these models because they are representative of code generation LLMs.
GPT-3 is used on popular code generation tools, such as GitHub Copilot~\cite{copilot} and ChatGPT~\cite{chatgpt}. CodeGen-2.5,  SantaCoder, StarCoder, and WizardCoder are open-source top-performing code LLMs~\cite{paperswithcode, luo2023wizardcoder}.

For each model, we generated \textbf{20} codes with a maximum of \textit{t} new tokens for each prompt. To choose a suitable value of maximum numbers of tokens \textit{t}, we calculated the size of canonical solutions for the \textsc{HumanEval}'s problems~\cite{HumanEval}. We found that the average solution has 54 tokens (maximum of 240 tokens). Hence, we asked each LLM to generate \textbf{512} new tokens (\ie  \textit{t} is around $10\times$ the average canonical solution's length).  
% We used the HuggingFace interface for the open source models. For GPT-3.5-Turbo, we used the OpenAI API to generate the code. 
Then, we calculated the \code{pass@k} metric by running the test cases for each output. 
% and the half-precision
\subsubsection*{Computing pass@k}
Code LLMs are commonly evaluated using \texttt{pass@k}~\cite{HumanEval, kulal2019spoc}. This metric estimates the probability that \textit{at least one} out of $k$ generated samples is correct (\ie passes all the prompt's test cases). This metric is computed by generating $n$ samples per prompt ($n \geq k$), counting the number of  samples  $c$  that are  correct ($c \leq n$), and calculating the unbiased estimator from Kulal \etal \cite{kulal2019spoc}:

\begin{equation}
    pass@k = \mathbb{E}_{prompts}\left[1- \frac{\binom {n-c}k}{\binom nk} \right]
\end{equation}

We set $k$ to 1, 3, and 10   and generated $n=20$ outputs for each prompt. We used temperature \textbf{1.0} for GPT-3.5-Turbo \cite{brown2020language} for all \code{pass@k}, \textbf{0.2} for \code{pass@1} and \textbf{0.6} for \code{pass@3} and \code{pass@10} for CodeGen model \cite{nijkamp2023codegen2}, and \textbf{0.2} for \code{pass@1} and \textbf{0.8} for  \code{pass@3} and \code{pass@10}  for the other open source models.
We chose these temperatures as these were the ones that were reported in the models' corresponding papers. 
In this evaluation, we compared the models' \code{pass@k}  when provided with the \textit{original} prompt and its \textit{fixed} versions.
% This way, for example, if there are 9 fixed prompts for an issue $I_n$, we compare the \code{pass@k} with the 9 original  and the 9 fixed prompts.

\begin{table*}[!ht]
\centering
\caption{Characteristics of the Studied Benchmarks}
\label{tab:dataset-info}
% \footnotesize
% \vspace{-10pt}
\setlength{\aboverulesep}{0pt}
\setlength{\belowrulesep}{0pt}
\setlength{\extrarowheight}{.75ex}
\setlength\tabcolsep{2pt} % default value: 6pt
\rowcolors{3}{gray!5}{white}
% \begin{tabular}{lP{13mm}P{57mm}ccP{32mm}}
\begin{tabular}{lP{13mm}P{75mm}cP{30mm}}
\toprule
\rowcolor[HTML]{efefef} \textbf{Name} & \textbf{\# Prompts} & \textbf{Target Code Language} &  \textbf{Usage Scenario} & \textbf{Contextual Dependency Complexity} \\
\midrule
CoderEval~\cite{2023arXiv230200288Y} & 460 &  Python, Java & Pragmatic Code Generation & self-contained \\
HumanEval~\cite{HumanEval} & 164 &  Python & Code Exercise & slib-runnable \\
JigsawDataset~\cite{jain2021jigsaw} & 87  & Python  & Public Library & \\
MBPP~\cite{MBPP} & 974 & Python & Code Exercise & self-contained\\
MXEVAL~\cite{MBXP}  & 16,171 & C\#, C++, Go, Java, JavaScript, Kotlin,      Perl, PHP, Python, Ruby, Scala, Swift, TypeScript & \begin{tabular}[l]{@{}l@{}}Code/Math Exercises\end{tabular} & slib-runnable, self-contained \\
\hspace{20pt}MBPP & 8,588 & \textit{All of the 13 languages listed above} & Code Exercises & slib-runnable, self-contained \\
\hspace{20pt}HumanEval & 1,934 & \textit{All except C++} & Code Exercises & slib-runnable \\
\hspace{20pt}MathQA & 5,649 & \textit{Only Python, Java, and JavaScript} & Math Exercises & slib-runnable, self-contained \\
NumpyEval~\cite{zan2022cert} & 101 & Python  & Single Public Library & plib-runnable\\
ODEX~\cite{wang2023executionbased} & 945 & Python & Open Domain & self-contained \\
PandasEval~\cite{zan2022cert} & 101 & Python  & Single Public Library & plib-runnable\\
TorchDataEval~\cite{zan2022language} & 50 & Python & Private Library & plib-runnable \\
\bottomrule

\end{tabular}
\end{table*}

\subsection{RQ4: Contamination Issues}

In this question, we checked the contamination issue of the widely used HumanEval's Python version benchmark \cite{HumanEval}. We used this benchmark because it is the only one that has the canonical solution, \ie it is at a higher risk of the contamination issue~\cite{sallou2023breaking}. 
To answer RQ4, we ran  NiCad (Automated Detection of Near-Miss Intentional Clones), a state-of-the-art code clone detection tool \cite{nicad}, on all code generated by each model. NiCad can detect different types of clones, including \textit{Type-1}, \textit{Type-2}, and \textit{Type-3} clones. 
%It uses a combination of text-based and tree-based approaches with support for further flexible source code normalization and transformation features using TXL \cite{cordy2006txl}. The tool provides a range of options for configuring the detection process, including parameters for defining the minimum size of clones and the level of similarity required to consider two code fragments as clones.

% To detect potential contamination issues, 

We first removed all the prompt's comments (docstring and in-line) from both the canonical solutions and the generated code. Then, we used NiCad cross-clone detection mechanism, which can find clones between two systems. In our case, the first system is the source codes from canonical solutions, and the second is the generated codes from the models. {We compared the canonical solutions not only with the generated codes from the \textit{original} prompts but also with the generated codes from the modified prompts that fix \textit{all} issues in RQ3. As the prompts were modified by us, they would not be a part of the training set, though they are similar to the original prompts. Hence, comparing the results with the original and the modified can provide us with more insights into the data contamination issue. }

We configured NiCad \cite{nicad} to find clones in the function labels, as the HumanEval dataset consists of prompts for completing functions. For Type-2 and Type-3 clone detection, we kept the default maximum difference threshold to 30\%. We configured the minimum line number of clones based on the size of the canonical solution line number.
That is, the \textit{minimum} number of lines is set to be half of the number of lines in the canonical solution.
It is worth highlighting that since NiCad can detect clones with at least 5 lines, we kept the threshold set to 5 in case the canonical solution had less than five lines. 
% The \textit{maximum} line number of a clone is set to the default value, \ie 2,500. 

Similar to a prior work~\cite{yang2024unveiling}, we used code clones as a means to identify cases where the generated code is identical (\ie a clone) to the solution. If a code clone is detected, the model likely has memorized the solution. Hence, to identify potential contamination issues,  we computed the percentage of different types of clones, including clones from the original and the modified prompts in the previous research questions. Notably, the Type-2 clone results from NiCad include Type-1 and Type-3 clone results include Type-1 and Type-2, and the result is kept as it is.

\subsection{RQ5: Quality Assessment of Developers' Prompts}

To investigate whether the quality issues observed in benchmark datasets (RQ2) are similar to the ones observed in real prompts, we have collected from the DevGPT dataset \textbf{3,995} publicly shared unique ChatGPT conversation links that are mentioned in code comments, commits, pull requests, discussions on GitHub, and threads on HackerNews~\cite{devgpt}. 
Next, we discarded 217 links that were no longer accessible.
% We kept 3,778 links that are still accessible. 
Then, we manually inspected each conversation to keep only those in which developers asked for code and  ChatGPT generated one or more code snippets according to the developers' prompts. That is, we discarded conversations in which ChatGPT did not generate code and/or the generated code was not in Java or Python. 
As a result, we had \textbf{371} ChatGPT conversation links that had generated Python/Java code. Two authors (the same ones from RQ2 \& RQ3) did the open coding of these 371 conversations. This open coding process  employed the same methodology as RQ2 (\shortref{subsec:RQ2Methodology}). Then, a senior author, who has more than ten years of experience, resolved any discrepancies. The Cohen's Kappa score to measure our inter-rater agreement was \textbf{0.60}, which indicates \textit{substantial agreement} \cite{mchugh2012interrater}.
% with an inter-rater agreement of 0.60, which indicates substantial agreement \cite{mchugh2012interrater}. 
% A senior author checked the discrepancies, and finally, 

\section{Results}

This section presents the results for each RQ.

\subsection{RQ1: Code Generation Benchmarks Comparison}\label{sec:ResultsRQ1}

As shown in Table \ref{tab:dataset-info}, the studied benchmarks have an average of \textbf{516} prompts per  language. In terms of \textit{supported programming languages}, we found that \textbf{all} benchmarks included prompts to generate Python code, and only \textbf{2} out of \textbf{9} benchmarks included other languages besides Python.  
The \textsc{MXEVAL}~\cite{MBXP} is a benchmark that extends three other benchmarks (\textsc{MathQA}~\cite{MathQA}, \textsc{MBPP}~\cite{MBPP}, and \textsc{HumanEval}~\cite{HumanEval}) to offer support to other \textbf{12}  languages besides Python. \textsc{CoderEval} is a benchmark created by mining Python and Java projects on GitHub and contains 230 prompts for each language. 

% Although programming languages can have multiple paradigms (\eg Python supports structured and object-oriented programming), the studied benchmarks mainly cover \textit{structured}, \textit{procedural}, and \textit{declarative} paradigms.
The benchmarks  did not have a variety of \textit{use scenarios}; their prompts were mostly crafted from coding exercises. Among these benchmarks, \textsc{CoderEval} \cite{CoderEval} and \textsc{ODEX} \cite{wang2023executionbased} cover problems from more diverse use cases; 
they had prompts that were based on GitHub repositories and StackOverflow questions which are more similar to real use cases.
% they had prompts whose solution would require using public libraries. 

In terms of \textit{contextual dependency complexity}, the benchmarks were mostly \textit{self-contained}, \textit{slib-runnable}, and \textit{plib-runnable}. This means the structure of the problem described in the prompt is simple, \ie it does not take context from different files under a project.

\begin{resultbox}
\textbf{{RQ1 Summary of Findings:}}
\begin{itemize}[leftmargin=*,label=-]
    \item {\textbf{Python} is the most supported language. Only \textbf{2} (out of \textbf{9}) benchmarks supported other languages besides Python.}
    % \item {The studied benchmarks mainly cover the \textit{\textbf{structured}}, \textit{\textbf{procedural}}, and \textit{\textbf{declarative}} paradigms.}
    % \item {The studied benchmarks had prompts whose solution would mostly require \textbf{built-in classes}. Only \textbf{3} benchmarks had prompts that would require using public libraries to solve the problem.} 
    \item Most benchmarks (\textbf{6} out of \textbf{9}) had prompts whose solution would mostly require \textbf{built-in classes}.
\end{itemize}
\end{resultbox}

\subsection{RQ2: Benchmark Quality Evaluation}

\begin{table*}[!ht]
\setlength{\aboverulesep}{0pt}
\setlength{\belowrulesep}{0pt}
\setlength{\extrarowheight}{.75ex}
\setlength\tabcolsep{1.5pt} % default value: 6pt
\caption{Quality Issues in each Benchmark (the percentages are the percent prompts with the issue in the benchmark).}
\label{tab:rq3_result} %\footnotesize\vspace{-10pt}
\rowcolors{3}{gray!5}{white}
\begin{tabular}{cp{5.8cm}cp{10cm}}
\toprule
\rowcolor[HTML]{efefef} \begin{tabular}[c]{@{}c@{}} \textbf{Type} \end{tabular}& \begin{tabular}[c]{@{}c@{}} \textbf{Quality Issue} \end{tabular}                                                       & \begin{tabular}[c]{@{}c@{}} \textbf{\# Prompts} \end{tabular} & \begin{tabular}[l]{@{}p{32mm}p{34mm}p{32mm}@{}} \textbf{\ Benchmark}\hfill\textbf{\%\ \ \ } & \textbf{\ Benchmark}\hfill\textbf{\%\ \ \ } & \textbf{\ Benchmark}\hfill\textbf{\%\ \ \ } \end{tabular}    \\ \midrule
% \parbox[t]{1.8mm}{\multirow{5}{*}{\rotatebox[origin=c]{90}{\textbf{Intent}}}}
% Intent
& \textbf{\issue{1}: Function/method's name mismatches with its intent} 	& 1,321 	& \begin{tabular}[l]{p{32mm}p{34mm}p{32mm}} \textsc{CoderEval}\hfill 0.6\% & \textsc{HumanEval}\hfill 9.1\% & \textsc{MBPP}\hfill 3.1\% \\ \textsc{MXEVAL\textsubscript{HumanEval}}\hfill 8.8\% & \textsc{MXEVAL\textsubscript{MBPP}}\hfill 3.2\% & \textsc{MXEVAL\textsubscript{MathQA}}\hfill 100.0\% \\ \textsc{NumpyEval}\hfill 1.1\% & \textsc{ODEX}\hfill 100.0\% & \textsc{TorchDataEval}\hfill 0.4\% \end{tabular} \\
\cellcolor{gray!00} & \textbf{\issue{2}: Spelling and grammatical errors} 	& 303 	& \begin{tabular}[l]{p{32mm}p{34mm}p{32mm}} \textsc{CoderEval}\hfill 8.2\% & \textsc{HumanEval}\hfill 17.4\% & \textsc{JigsawDataset}\hfill 9.6\% \\ \textsc{MBPP}\hfill 2.7\% & \textsc{MXEVAL\textsubscript{HumanEval}}\hfill 17.6\% & \textsc{MXEVAL\textsubscript{MBPP}}\hfill 4.7\% \\ \textsc{MXEVAL\textsubscript{MathQA}}\hfill 6.7\% & \textsc{NumpyEval}\hfill 10.2\% & \textsc{ODEX}\hfill 4.2\% \\ \textsc{PandasEval}\hfill 13.6\% & \textsc{TorchDataEval}\hfill 20.7\% &\end{tabular} \\
\cellcolor{gray!00}  & \textbf{\issue{3}: The prompt description is unclear} 	& 124 	& \begin{tabular}[l]{p{32mm}p{34mm}p{32mm}} \textsc{CoderEval}\hfill 2.9\% & \textsc{HumanEval}\hfill 1.5\% & \textsc{MXEVAL\textsubscript{HumanEval}}\hfill 0.4\% \\ \textsc{MXEVAL\textsubscript{MBPP}}\hfill 6.3\% & \textsc{MXEVAL\textsubscript{MathQA}}\hfill 4.2\% & \textsc{NumpyEval}\hfill 5.7\% \\ \textsc{ODEX}\hfill 1.9\% & \textsc{PandasEval}\hfill 3.4\% & \textsc{TorchDataEval}\hfill 2.6\% \end{tabular} \\
\cellcolor{gray!00}& \textbf{\issue{4}: Partial or incomplete sentence} 	& 132 	& \begin{tabular}[l]{p{32mm}p{34mm}p{32mm}} \textsc{MXEVAL\textsubscript{MathQA}}\hfill 13.3\% & \textsc{ODEX}\hfill 0.4\% & \end{tabular} \\
\cellcolor{gray!00} \parbox[t]{5mm}{\multirow{-12}{*}{\rotatebox[origin=c]{90}{\textbf{Intent}}}} & \textbf{\issue{5}: Incorrect input/output pair example} 	& 15 	& \begin{tabular}[l]{p{32mm}p{34mm}p{32mm}} \textsc{HumanEval}\hfill 1.5\% & \textsc{MXEVAL\textsubscript{HumanEval}}\hfill 1.5\% & \textsc{MXEVAL\textsubscript{MBPP}}\hfill 1.1\% \end{tabular} \\ 
\hline
% Formatting
\cellcolor{gray!00} & \textbf{\issue{6}: JavaDoc/docstring has formatting issues} 	& 1,835 	& \begin{tabular}[l]{p{32mm}p{34mm}p{32mm}} \textsc{HumanEval}\hfill 22.0\% & \textsc{MXEVAL\textsubscript{HumanEval}}\hfill 81.7\% & \textsc{MXEVAL\textsubscript{MBPP}}\hfill 89.5\% \\ \textsc{MXEVAL\textsubscript{MathQA}}\hfill 89.2\% & \textsc{NumpyEval}\hfill 1.1\% & \textsc{PandasEval}\hfill 2.3\% \\ \textsc{TorchDataEval}\hfill 1.1\% & &\end{tabular} \\
\cellcolor{gray!00} & \textbf{\issue{7}: Not using JavaDoc (Java) or docstring (Python) on the prompt} 	& 439 	& \begin{tabular}[l]{p{32mm}p{34mm}p{32mm}} \textsc{HumanEval}\hfill 2.3\% & \textsc{MXEVAL\textsubscript{HumanEval}}\hfill 0.8\% & \textsc{NumpyEval}\hfill 97.7\% \\ \textsc{PandasEval}\hfill 94.3\% & \textsc{TorchDataEval}\hfill 98.1\% & \end{tabular} \\
\cellcolor{gray!00} \parbox[t]{5mm}{\multirow{-6}{*}{\rotatebox[origin=c]{90}{\textbf{Formatting}}}} & \textbf{\issue{8}: Inconsistent prompt style} 	& 311 	& \begin{tabular}[l]{p{32mm}p{34mm}p{32mm}} \textsc{HumanEval}\hfill 88.6\% & \textsc{MXEVAL\textsubscript{HumanEval}}\hfill 66.4\% & \textsc{MXEVAL\textsubscript{MBPP}}\hfill 1.4\% \\ \textsc{NumpyEval}\hfill 3.4\% & \textsc{PandasEval}\hfill 2.3\% & \textsc{TorchDataEval}\hfill 1.5\% \end{tabular} \\
\hline
% Noise
\cellcolor{gray!00} & \textbf{\issue{9}: Interrogation questions in the prompt} 	& 152 	& \begin{tabular}[l]{p{32mm}p{34mm}p{32mm}} \textsc{MXEVAL\textsubscript{MathQA}}\hfill 0.1\% & \textsc{NumpyEval}\hfill 44.3\% & \textsc{ODEX}\hfill 0.4\% \\ \textsc{PandasEval}\hfill 36.4\% & \textsc{TorchDataEval}\hfill 29.3\% &\end{tabular} \\
\cellcolor{gray!00} \parbox[t]{5mm}{\multirow{-3}{*}{\rotatebox[origin=c]{90}{\textbf{Noise}}}}& \textbf{\issue{10}: URL or reference in the comment} 	& 18 	& \begin{tabular}[l]{p{32mm}p{34mm}p{32mm}} \textsc{MBPP}\hfill 6.1\% & \textsc{MXEVAL\textsubscript{MBPP}}\hfill 0.3\% & \end{tabular} \\
\bottomrule
\end{tabular}
\end{table*}

\begin{figure}[!ht]
    \centering
    \includegraphics[width=.6\linewidth]{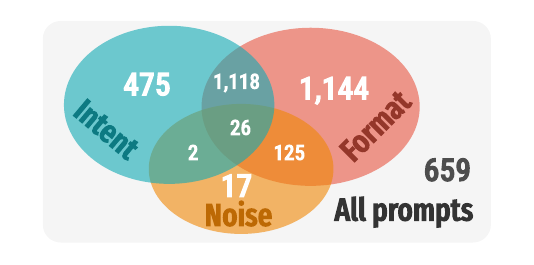}\vspace{-10pt}
    \caption{Distribution of quality issues types}
    \label{fig:venn}
\end{figure}

From our open coding of benchmarks' prompts, we found \textbf{10} quality issues that can be classified into \textbf{3} main categories. 
Figure~\ref{fig:venn} shows the quality issue types we found and their counts, while Table~\ref{tab:rq3_result} enumerates the quality issue types we found. Most issues were related to the prompt's \textit{format} and \textit{intent}; there were \textbf{2,413} (\textbf{68\%}) prompts improperly formatted and \textbf{1,621} (\textbf{53\%}) with issues that may affect the LLM's ability to understand the prompt.   Only \textbf{659} (18\%) of the analyzed prompts did not have any issues with them.
% We also did not find prompts with \textit{very short sentences}, \textit{self-admitted technical debt},  \textit{automatically generated comments}, and \textit{comments that include PIIs}.
% some issues we initially targeted, like very short sentences, Admitted Technical Debt (SATD), auto-generated code/comments, and non-declarative sentences in the prompts.

% P2: Very short sentence
% P4: Self-Admitted Technical Debt
% P5: Automatically generated code/comments
% P14: Comment includes PII(author name, email address, etc.)

\subsubsection*{--\textbf{\underline{Intent-related issues}}}
This category refers to quality issues that can affect the LLM's ability to understand the intent (\ie purpose or goal)  behind the prompt. 
We noticed that \textbf{\textit{all}} benchmarks had at least one prompt with \textsl{\textbf{spelling and grammatical errors}} in them (\textbf{\issue{2}}). However, most of the prompts in these benchmarks were grammatically correct; only between \textbf{2.7\%} to \textbf{20.7\%} of them had spelling/grammatical errors.
We also found that these benchmarks had prompts whose \textsl{\textbf{function/method's name does not match the intention described in the prompt}} (\textbf{\issue{1}}). It means the benchmark's developers used names that do not make the intended functionality clear. For example,  all prompts in \textsc{ODEX} have this issue as the prompts'  function name is in this format \snippetPython{f_Prompt_ID}. Similarly, all prompts' functions in \textsc{MXEVAL\textsubscript{MathQA}} are named \snippetPython{problem}. We also found that two benchmarks had prompts with \textbf{\textsl{partial/incomplete sentences}} (\textbf{\issue{4}}). Moreover, 
\textsc{MXEVAL} and \textsc{HumanEval} have \textbf{\textsl{incorrect sample input-output pairs}} (\textbf{\issue{5}});  our analysis showed that $\approx$1\% of their prompts are wrong. Prompts with incorrect examples of input/output pairs give inaccurate contextual information to the model. 
For instance, the HumanEval's prompt in Listing \ref{lst:P16_example_wrongio}  should have \codePython{None} instead of an empty line (line 8).

\begin{listing}[H]
{
\begin{PythonSourceCode*}{label=\textcolor{black}{\sf\tiny{\textbf{Dataset}: HumanEval~\cite{HumanEval}. \textbf{Prompt ID}: 12}}}
from typing import List, Optional
def longest(strings: List[str]) -> Optional[str]:
  """ 
  Out of list of strings, return the longest one. Return the first one in case of
  multiple strings of the same length. 
  Return None in case the input list is empty.
  >>> longest([])
    
  >>> longest(['a', 'b', 'c'])
  'a'
  >>> longest(['a', 'bb', 'ccc'])
  'ccc'
  """
\end{PythonSourceCode*}
}
% \vspace{-10pt}
\caption{Example of an incorrect input-output pair.}\label{lst:P16_example_wrongio}
\end{listing}

\subsubsection*{--\textbf{\underline{Format-related issues}}}
This category refers to problems related to how prompts are formatted.
We found that \textbf{7} benchmarks \textbf{\textsl{did not properly use Javadocs/docstrings to express the function/method's intent}} (\textbf{\issue{6}}). This was especially pervasive on the \textsc{MXEVAL} benchmark; over \textbf{81\%} of its prompts did not use proper Javadocs/docstrings. Moreover,   \textbf{439} prompts  were \textbf{\textsl{using single/multi-line comments to describe the intended behavior instead of using docstrings or Javadocs}} (\textbf{\issue{7}}). We also found \textbf{\textsl{inconsistent formatting in the benchmarks}}, \ie style inconsistencies in them (\textbf{\issue{8}}). For example, we observed  Python benchmarks in which some prompts included type annotations, but others did not.  
% To do so, we first created a dataset profile for each dataset, which documents the typical format of each prompt, \eg whether it includes Python's type annotations, which docstring formatting style is used (\eg reST, Google, \etc), whether it has input/output examples on the description, \etc~
% Subsequently, we inspected each prompt in the dataset to verify whether it deviated from the expected format.  .  

\begin{comment}
For example, if we compare the following sample from HumanEval \cite{HumanEval} in Listing \ref{lst:inconsistant} with the sample in Listing \ref{lst:P16_example_wrongio}, we can see that they are using different quote styles (\ie double quote is the standard \cite{Docstring}), the number of examples input and output are different, and one sample is having type hint while another one is not.

\input{Codes/inconsitant}
\end{comment}

\subsubsection*{--\textbf{\underline{Noise-related issues}}}
This category refer to cases where prompts contain unnecessary tokens (\textit{noise}). 
We found \textbf{152} (\textbf{4\%}) prompts with \textsl{\textbf{confusion questions}}, \eg \textit{``Is there a nice Pythonic way to do this?''} (\textbf{\issue{9}}). Another noise-related issue found was \textsl{\textbf{URLs in the prompt}} (\textbf{\issue{10}}), which do not carry meaningful information for the model.

\begin{resultbox}
\textbf{{RQ2 Summary of Findings:}}
\begin{itemize}[leftmargin=*,label=-]
    \item \textbf{2907} (\textbf{82\%}) of studied prompts had at least one quality issue in them. 
    \item \textbf{\textsl{Javadoc/docstring formatting issues}}, \textbf{\textsl{function/method's name mismatching its intent}}, and \textbf{\textsl{spelling/grammatical errors}}, were the three most common quality issues.
    % \item All benchmarks had \textit{\textbf{at least one}} prompt with spelling and grammatical errors.
\end{itemize}
\end{resultbox}

\subsection{RQ3: Impact on Performance}

We ran five LLMs with the \textit{original} prompt and \textit{fixed} prompts. 
To better understand how each quality issue may affect an LLM's performance, we created prompts that fixed \textit{one} issue at a time and prompts that fixed \textit{all} issues. The \green{green} cells in Tables \ref{tab:rq4_result_java} \& \ref{tab:rq4_result_python} highlight the case in which the \texttt{pass@k} of the \textit{fixed} prompt was higher than the \textit{original} prompt.

\begin{table}[!ht]
\centering
\caption{Pass@k Comparison (original \textit{vs.} fixed prompts -- Java)}
\label{tab:rq4_result_java} % \footnotesize\vspace{-10pt}
\scriptsize
\setlength{\aboverulesep}{0pt}
\setlength{\belowrulesep}{0pt}
\setlength{\extrarowheight}{.75ex}
\setlength\tabcolsep{1.8pt} % default value: 6pt
\rowcolors{3}{gray!5}{white}
\begin{tabular}{@{}ccccccccc@{}}
\toprule
\multirow{2}{*}{\textbf{Model}}  & \multirow{2}{*}{\textbf{\begin{tabular}[c]{@{}c@{}}Fixed\\ Issue\end{tabular}}} & \multirow{2}{*}{\textbf{Total}} & \multicolumn{3}{c}{\textbf{Modified prompt with fixes}} & \multicolumn{3}{c}{\textbf{Original prompt}} \\
 &  &  &  \textbf{pass@1} & \textbf{pass@3} & \textbf{pass@10} & \textbf{pass@1} & \textbf{pass@3} & \textbf{pass@10} \\
 \midrule
\textbf{CodeGen} & \issue{1} & 6 & 0.000 & 0.000 & 0.000 & 0.100 & 0.302 & 0.639 \\
\textbf{CodeGen} & \issue{2} & 22 & 0.155 & \green{0.488} & \green{0.795} & 0.164 & 0.483 & 0.789 \\
\textbf{CodeGen} & \issue{3} & 1 & 0.000 & 0.000 & 0.000 & 0.000 & 0.000 & 0.000 \\
\textbf{CodeGen} & \issue{5} & 1 & \green{0.600} & 0.509 & 0.957 & 0.500 & 0.681 & 0.995 \\
\textbf{CodeGen} & \issue{6} & 161 & 0.157 & 0.444 & 0.768 & 0.219 & 0.494 & 0.807 \\
\textbf{CodeGen} & \issue{8} & 67 & 0.153 & 0.305 & 0.494 & 0.184 & 0.448 & 0.751 \\
\textbf{CodeGen} & All & 161 & 0.150 & 0.375 & 0.620 & 0.219 & 0.494 & 0.807 \\\hline
\textbf{SantaCoder} & \issue{1}  & 6   & 0.000 & 0.000 & 0.000 & 0.758 & 0.819 & 0.917 \\
\textbf{SantaCoder} & \issue{2}  & 22  & 0.675 & 0.823 & 0.948 & 0.709 & 0.850 & 0.975 \\
\textbf{SantaCoder} & \issue{3}  & 1   & 0.050 & 0.150 & 0.500 & 0.050 & 0.150 & 0.500 \\
\textbf{SantaCoder} & \issue{5}  & 1   & \green{0.200} & 0.681 & 0.995 & 0.150 & 0.855 & 1.000 \\
\textbf{SantaCoder} & \issue{6}  & 161 & 0.609 & 0.781 & 0.938 & 0.690 & 0.848 & 0.961 \\
\textbf{SantaCoder} & \issue{8} & 67  & 0.410 & 0.494 & 0.555 & 0.596 & 0.800 & 0.962 \\
\textbf{SantaCoder} & All & 161 & 0.532 & 0.646 & 0.738 & 0.690 & 0.848 & 0.961\\ \hline
\textbf{StarCoder} & \issue{1} & 6 & 0.000 & 0.000 & 0.000 & 0.600 & 0.725 & 0.831 \\
\textbf{StarCoder} & \issue{2} & 22 & \green{0.880} & 0.853 & 0.941 & 0.866 & 0.901 & 0.972 \\
\textbf{StarCoder} & \issue{3} & 1 & 0.000 & 0.150 & 0.500 & 0.000 & 0.150 & 0.500 \\
\textbf{StarCoder} & \issue{5} & 1 & 0.800 & 0.926 & 1.000 & 0.900 & 0.982 & 1.000 \\
\textbf{StarCoder} & \issue{6} & 161 & 0.717 & 0.845 & \green{0.947} & 0.755 & 0.869 & 0.939 \\
\textbf{StarCoder} & \issue{8} & 67 & 0.432 & 0.508 & 0.562 & 0.742 & 0.847 & 0.958 \\
\textbf{StarCoder} & All & 161 & 0.574 & 0.684 & 0.744 & 0.755 & 0.869 & 0.939 \\\hline
\textbf{WizardCoder} & \issue{1}  & 6   & 0.000 & 0.000 & 0.000 & 0.758 & 0.741 & 0.826 \\
\textbf{WizardCoder} & \issue{2}  & 22  & \green{0.680} & \green{0.803} & \green{0.903} & 0.659 & 0.787 & 0.893 \\
\textbf{WizardCoder} & \issue{3}  & 1   & 0.000 & 0.000 & 0.000 & 0.000 & 0.000 & 0.000 \\
\textbf{WizardCoder} & \issue{5}  & 1   & 0.950 & \green{0.926} & \green{1.000} & 0.950 & 0.601 & 0.984 \\
\textbf{WizardCoder} & \issue{6}  & 161 & 0.641 & 0.798 & 0.898 & 0.683 & 0.827 & 0.919 \\
\textbf{WizardCoder} & \issue{8} & 67  & 0.403 & 0.494 & 0.530 & 0.617 & 0.761 & 0.905 \\
\textbf{WizardCoder} & All & 161 & 0.534 & 0.665 & 0.724 & 0.683 & 0.827 & 0.919 \\ \hline
\textbf{GPT-3.5} & \issue{1} & 6 & 0.000 & 0.000 & 0.000 & 0.875 & 0.976 & 1.000 \\
\textbf{GPT-3.5} & \issue{2} & 22 & 0.883 & 0.950 & 0.952 & 0.900 & 0.952 & 0.952 \\
\textbf{GPT-3.5} & \issue{3} & 1 & 0.000 & 0.000 & 0.000 & 0.000 & 0.000 & 0.000 \\
\textbf{GPT-3.5} & \issue{5} & 1 & 0.550 & 0.926 & 1.000 & 0.750 & 0.991 & 1.000 \\
\textbf{GPT-3.5} & \issue{6} & 161 & 0.860 & 0.941 & 0.956 & 0.876 & 0.945 & 0.961 \\
\textbf{GPT-3.5} & \issue{8} & 67 & 0.515 & 0.547 & 0.552 & 0.813 & 0.913 & 0.944 \\
\textbf{GPT-3.5} & All & 161 & 0.713 & 0.761 & 0.770 & 0.876 & 0.945 & 0.961 \\
\bottomrule
\end{tabular}
\end{table}

\begin{table}[!ht]
\centering
\caption{Pass@k Comparison (original \textit{vs.} fixed prompts -- Python)}
\label{tab:rq4_result_python} % \footnotesize\vspace{-10pt}
\scriptsize
\setlength{\aboverulesep}{0pt}
\setlength{\belowrulesep}{0pt}
\setlength{\extrarowheight}{.75ex}
\setlength\tabcolsep{1.5pt} % default value: 6pt
\rowcolors{3}{gray!5}{white}
\begin{tabular}{ccccccccc}
\toprule
\multirow{2}{*}{\textbf{Model}}  & \multirow{2}{*}{\textbf{\begin{tabular}[c]{@{}c@{}}Fixed\\ Issue\end{tabular}}} & \multirow{2}{*}{\textbf{Total}} & \multicolumn{3}{c}{\textbf{Modified prompt with fixes}} & \multicolumn{3}{c}{\textbf{Original prompt}} \\
 &  &  &  \textbf{pass@1} & \textbf{pass@3} & \textbf{pass@10} & \textbf{pass@1} & \textbf{pass@3} & \textbf{pass@10} \\
 \midrule
\textbf{CodeGen}  & \issue{2} & 27 & 0.093 & 0.172	 & 0.341 & 0.104 & 0.190& 0.363 \\
% \textbf{CodeGen}  & I4 & 1 & 0.000 & 0.000 & 0.000 & 0.000 & 0.000 & 0.000 \\
\textbf{CodeGen}  & \issue{1} & 7 & 0.000 & 0.000 & 0.000 & 0.014 &0.138 & 0.251\\
\textbf{CodeGen}  & \issue{3} & 1 & 0.000 & 0.000 & 0.000 & 0.000 & 0.000 & 0.000 \\
\textbf{CodeGen}  & \issue{6} & 164 & 0.045 & 0.187& 0.352 & 0.066 &0.225 & 0.423 \\
\textbf{CodeGen}  & \issue{8} & 147 & 0.057 & 0.121 & 0.228 & 0.058 & 0.121 & 0.237 \\
\textbf{CodeGen}  & All & 164 & 0.039 &0.155& 0.297 & 0.066 & 0.209 & 0.394 \\ \hline
\textbf{SantaCoder}  & \issue{2}  & 27  & 0.000 & 0.000 & 0.000 & 0.000 & 0.006 & 0.019 \\
% \textbf{SantaCoder}  & I4  & 1   & 0.000 & 0.000 & 0.000 & 0.000 & 0.000 & 0.000 \\
\textbf{SantaCoder}  & \issue{1}  & 7   & 0.000 & 0.000 & 0.000 & 0.000 & 0.000 & 0.000 \\
\textbf{SantaCoder}  & \issue{3}  & 1   & 0.000 & 0.000 & 0.000 & 0.000 & 0.000 & 0.000 \\
\textbf{SantaCoder}  & \issue{6}  & 164 & 0.002 & \green{0.022} & \green{0.063} & 0.002 & 0.016 & 0.045 \\
\textbf{SantaCoder}  & \issue{8} & 147 & 0.002 & \green{0.019} & \green{0.054} & 0.002 & 0.011 & 0.030 \\
\textbf{SantaCoder}  & All & 164 & \green{0.005} & \green{0.033} & \green{0.081} & 0.002 & 0.016 & 0.045 \\ \hline
\textbf{StarCoder}  & \issue{2} & 27 & 0.002 & 0.054 & \green{0.168} & 0.013 & 0.058 & 0.164 \\
% \textbf{StarCoder}  & I4 & 1 & 0.000 & 0.000 & 0.000 & 0.000 & 0.000 & 0.000 \\
\textbf{StarCoder}  & \issue{1} & 7 & 0.000 & 0.000 & 0.000 & 0.057 & 0.079 & 0.199 \\
\textbf{StarCoder}  & \issue{3} & 1 & 0.000 & 0.000 & 0.000 & 0.000 & 0.000 & 0.000 \\
\textbf{StarCoder}  & \issue{6} & 164 & 0.023 & \green{0.100} & \green{0.232} & 0.035 & 0.097 & 0.225 \\
\textbf{StarCoder}  & \issue{8} & 147 & \green{0.035} & \green{0.086} & 0.199 & 0.029 & 0.085 & 0.204 \\
\textbf{StarCoder}  & All & 164 & 0.025 & \green{0.105} & \green{0.246} & 0.035 & 0.097 & 0.225 \\
\hline

\textbf{WizardCoder}  & \issue{2}  & 27  & \green{0.013} & 0.011 & 0.037 & 0.006 & 0.038 & 0.121 \\
% \textbf{WizardCoder}  & I4  & 1   & 0.000 & \green{0.150} & \green{0.500} & 0.000 & 0.000 & 0.000 \\
\textbf{WizardCoder}  & \issue{1}  & 7   & 0.000 & 0.000 & 0.000 & 0.007 & 0.043 & 0.143 \\
\textbf{WizardCoder}  & \issue{3}  & 1   & 0.000 & 0.000 & 0.000 & 0.000 & 0.000 & 0.000 \\
\textbf{WizardCoder}  & \issue{6}  & 164 & 0.057 & 0.095 & 0.187 & 0.085 & 0.124 & 0.236 \\
\textbf{WizardCoder}  & \issue{8} & 147 & 0.063 & 0.095 & 0.172 & 0.085 & 0.124 & 0.236 \\
\textbf{WizardCoder}  & All & 164 & 0.005 & 0.033 & 0.081 & 0.002 & 0.016 & 0.045 \\ \hline

\textbf{GPT-3.5}  & \issue{2} & 27 & \green{0.235} & \green{0.450} & 0.638 & 0.219 & 0.422 & 0.639 \\
% \textbf{GPT-3.5}  & I4 & 1 & 0.050 & 0.150 & 0.500 & 0.750 & 0.991 & 1.000 \\
\textbf{GPT-3.5}  & \issue{1} & 7 & 0.000 & 0.000 & 0.000 & 0.393 & 0.500 & 0.565 \\
\textbf{GPT-3.5}  & \issue{3} & 1 & 0.000 & 0.000 & 0.000 & 0.000 & 0.000 & 0.000 \\
\textbf{GPT-3.5}  & \issue{6} & 164 & \green{0.274} & \green{0.458} & \green{0.639} & 0.249 & 0.402 & 0.549 \\
\textbf{GPT-3.5}  & \issue{8} & 147 & \green{0.295} & \green{0.459} & \green{0.630} & 0.275 & 0.441 & 0.590 \\
\textbf{GPT-3.5}  & All & 164 & \green{0.275} & \green{0.454} & \green{0.640} & 0.249 & 0.402 & 0.549 \\
\hline
\bottomrule
\end{tabular}
\end{table}

We found that after fixing spelling and grammatical issues (\textbf{\issue{2}}), the CodeGen,  and WizardCoder models, on average, performed better than the original Java prompts. Prompts with a correct JavaDoc and Docstring style (\textbf{\issue{6}}) tended to perform better than compared to the original prompts. Creating a consistent prompting style across the dataset is better for Python prompts from the GPT-3.5-Turbo model.

When fixing \textit{incorrect input/output pair examples} (\issue{5}) we noticed that the \code{pass@1} improved for CodeGen and StarCoder. 
While we observed cases where fixing one (or all) issues in a prompt increased a model's \code{pass@k}, there was not a consistent trend across models and languages.
Fixing all issues in a Python prompt increased the \code{pass@k} of  SantaCoder, StarCoder, and GPT-3.5-Turbo models.

\begin{resultbox}
\textbf{{RQ3 Summary of Findings:}}
\begin{itemize}[leftmargin=*,label=-]
    \item Fixing spelling and grammatical issues and having the standard JavaDoc and Docstring style can perform similarly to original prompts.
    \item Fixing different quality issues in a single prompt can provide better performance for Python code generation. 
\end{itemize}
\end{resultbox}

\subsection{RQ4: Test set Contamination}
We ran the NiCad \cite{nicad} tool to detect generated codes that are clones of the canonical solutions in the HumanEval dataset. 
We can not see any Type-1 and hardly see Type-2 clones in the SantaCoder, StarCoder, and WizardCoder after using this tool. These models use the Stack dataset \cite{Kocetkov2022TheStack}, and we checked if the original HumanEval dataset from OpenAI in this dataset using a tool provided by them\footnote{\url{https://huggingface.co/spaces/bigcode/in-the-stack}}. Our result and the tool confirm that \textsl{\textbf{there is no test set contamination issue for the HumanEval dataset for these models}}. The CodeGen-2.5 models are also trained with the Stack dataset \cite{Kocetkov2022TheStack}  and three other training sets. There are comparatively more Type-1 and Type-2 clones from the output of this model. This indicates that test set contamination issues are present in the CodeGen-2.5  model. 

The result for GPT-3.5 has more Type-1 and Type-2 clones than other models. As it is a closed source model, we cannot directly verify its training set to check whether it includes the HumanEval benchmark. However, given the  higher code clone incidence, this model may have HumanEval's solution in its training set, which can justify the high performance of this model. We can also see that modified prompts generate fewer or equal numbers of clones than the original prompts. 
% This can be a potential solution for data contamination issues for old evaluation benchmarks released before the knowledge cut-off. 

\begin{table}[!ht]
\setlength{\aboverulesep}{0pt}
\setlength{\belowrulesep}{0pt}
\setlength{\extrarowheight}{.75ex}
\setlength\tabcolsep{2pt} % default value: 6pt
\caption{RQ5 results for each LLMs and temperature (\textbf{T}).}
\label{tab:code_clone}% \footnotesize\vspace{-10pt}
\scriptsize
\rowcolors{3}{gray!10}{white}
\begin{tabular}{lccccccc}
\toprule
   \multicolumn{1}{c}{}         &   \multicolumn{1}{c}{}     & \multicolumn{3}{c}{\textbf{Original}} & \multicolumn{3}{c}{\textbf{Modified}} \\
\midrule
\textbf{Model}     & \textbf{T} & \textbf{Type-1}   & \textbf{Type-2}  & \textbf{Type-3}  & \textbf{Type-1}   & \textbf{Type-2}  & \textbf{Type-3}  \\
CodeGen   & 0.2  & 1 (0.6\%)      & 2 (1.2\%)      & 10 (6.1\%)     & 0 (0.0\%)     & 2 (1.2\%)      & 3 (1.8\%)     \\
CodeGen   & 0.6  & 1 (0.6\%)      & 5 (3.0\%)      & 21 (12.8\%)     & 0 (0.0\%)      & 4 (2.4\%)     & 20 (12.20\%)    \\
SantaCoder   & 0.2  & 0 (0.0\%)      & 0 (0.0\%)      & 0 (0.0\%)     & 0 (0.0\%)     & 0 (0.0\%)      & 0 (0.0\%)     \\
SantaCoder   & 0.8  & 0 (0.0\%)      & 0 (0.0\%)      & 1 (0.6\%)     & 0 (0.0\%)      & 0 (0.0\%)     & 1 (0.6\%)    \\
StarCoder & 0.2  & 0 (0.0\%)       & 0 (0.0\%)      & 1 (0.6\%)       & 0 (0.0\%)       & 0 (0.0\%)      & 0 (0.0\%)     \\
StarCoder & 0.8  & 0 (0.0\%)        & 1 (0.6\%)      & 8 (4.9\%)      &2 (1.2\%)        & 4 (2.4\%)       & 10 (6.1\%)       \\
WizardCoder   & 0.2  & 0 (0.0\%)      & 0 (0.0\%)      & 1 (0.6\%)     & 0 (0.0\%)      & 0 (0.0\%)     & 1 (0.6\%)    \\
WizardCoder   & 0.8  & 0 (0.0\%)      & 1 (0.6\%)      & 5 (3.0\%)     & 0 (0.0\%)      & 0 (0.0\%)     & 4 (2.4\%)    \\
GPT-3.5   & 1.0   & 4 (2.4\%)        & 11 (6.7\%)      & 32 (19.5\%)       & 3 (1.8\%)        & 11 (6.7\%)       & 42 (25.6\%)    \\ 

\noalign{\global\arrayrulewidth=1pt}
  \arrayrulecolor{black}\bottomrule
\end{tabular}
\end{table}

\begin{resultbox}
\textbf{{RQ4 Summary of Findings:}}

 The dataset used for training the \textsc{CodeGen-2.5} and \textsc{GPT-3.5} model has a data contamination issue with the HumanEval dataset.
\end{resultbox}

\subsection{RQ5: Quality Assessment of Developers' Prompts}

We identified \textbf{11} quality issues
from \textbf{198} conversations 
between developers and ChatGPT (\ie 54\% of the conversations had \textit{at least one} quality issue). Similar to RQ2, these issues are classified into three categories:  

% We have analyzed the conversations availble with from the DevGPT \cite{devgpt} benchmarks 
\subsubsection*{--\textbf{\underline{Intent-related issues}}}

This category refers to the developers' intention to describe the task for the ChatGPT.

\begin{itemize}[leftmargin=*,label=-,noitemsep,topsep=0pt]
% \item \textbf{No code generated as response} To test a code generation
% tool’s capability, the prompt response should contain some
% code, not just plain text. However, in some cases, the code
% generation tools give step-by-step instructions or processes
% rather than code. There are 60 conversations where this
% happened.

    \item \textbf{Unclear prompt description:}
    When the description is ambiguous, confusing, and not precise, it may lead to a very different output than the intended one. It can be caused by a lack of detailed information or user mistakes. For \textbf{57} conversations, we observed that ChatGPT did not understand the prompt description and provided an output that did not fulfill what the developer asked. 
    
    \item \textbf{Spelling or grammatical error:}
    % We checked if the prompt was grammatically correct, with the proper spelling of each word. Spelling and grammatical errors can cause the code-generation tool to misinterpret the user's intention. 
    There were \textbf{53} developers' prompts which had a spelling or grammatical error.
    
    \item \textbf{Lack of enough context:}
    Many parameters and values, as well as the description of related classes and objects, might be necessary to generate complex code. Problematic prompts ask to generate complex code without specifying proper context. 
    We found \textbf{51} conversations that missed important context to solve the task described in the prompt.

    \item  \textbf{Very Short Sentence:}
    We  found \textbf{3} conversations with very short sentences (less than 3 words in length). 
    
    \item \textbf{Self Admitted Technical Debt (SATD)}
     We found \textbf{6} prompts that contained Self Admitted Technical Debt (SATD). 
    These are comments written by developers to indicate buggy, incomplete, or suboptimal codes (\eg \code{TODO})~\cite{potdar2014satd}.
    % from the developers were containing SATD.
\end{itemize}

\subsubsection*{--\textbf{\underline{Formatting-related issues}}}
This category pertains to the improper formatting done by the developers while describing a coding task to ChatGPT.

\begin{itemize}[leftmargin=*,label=-,noitemsep,topsep=0pt]
    \item \textbf{Not using standard JavaDoc or Docstring:}
    We checked if the prompts follow standard JavaDoc or Docstring for Java or Python, respectively. Providing input in a standard format should help the code generation tool better utilize it. We found that developers did not use proper JavaDoc or Docstring for \textbf{43} conversations. 

\item \textbf{Messy code snippet:}
    Codes in the prompts can be messy, with no or incorrect indentation and spacing. There were \textbf{14} prompts where the included code snippets were messy. 
\end{itemize}

\subsubsection*{--\textbf{\underline{Noise-related issues}}}
The prompt from the developers can contain unnecessary portions that may not be helpful in expressing the context.

\begin{itemize}[leftmargin=*,label=-,noitemsep,topsep=0pt]
    \item \textbf{URL Link or reference:}
    The prompt may contain a URL Link or reference to an external source. As some versions of ChatGPT may not be able to browse the Webpage using the link, the link will not add any additional information to the prompt. There were \textbf{15} prompts from the developers, which included URLs. 

    % \item \textbf{Question in the prompt}
    % Some prompts contain coding-related questions, but that are open-ended and mostly about how a function or library works. These are not useful prompts for code-generation tools. \textbf{53} prompts had this quality issue. 
\end{itemize}

\paragraph*{\textbf{Comparison with benchmarks' prompts}} In RQ2, we analyzed quality issues on prompts within benchmarks, and in this RQ we perform the same analysis over developers' prompts to ChatGPT. In our analyses, we found that intent-related and formatting-related quality issues are not the same in both prompt types. The developers' prompts included SATD and very short sentences. 
% Developers' prompts also have a quality issue with the presentation of the code in the prompt. T
Moreover,  RQ2 results showed cases that the benchmark prompts had \textit{incorrect} input/output pair examples, and the function name did  not match the intent. 
% The benchmark prompts had issues related with  adherence to a consistent prompting style.

\begin{resultbox}
\textbf{{RQ5 Findings:}}
\begin{itemize}[leftmargin=*,label=-]
    \item 54\% of the conversations from the developers with ChatGPT had at least one issue.
    \item Prompts from developers mostly lack enough context. 
\end{itemize}
\end{resultbox}

\section{Discussion}

 \paragraph*{\textbf{Benchmarks lack diversity}}
Our RQ1 findings indicate that the benchmarks mainly focus on Python and have small code exercises. According to a recent survey with about 67,000 professional developers from Stack Overflow \cite{so2023survey}, the most popular language is JavaScript. At the same time, Typescript is close to Python, and Java, C\#, and C++ are not that behind. This indicates that we need to expand the benchmark dataset beyond Python. It is also noticeable from our findings that the benchmarks are not that complex. At the same time, real-world software can have thousands of lines of code intra and inter-dependencies with local and public libraries \cite{meidani2022towards}. Hence, the current benchmarks do not mimic real-world scenarios. That is also indicated by a recent study \cite{siddiq2023empirical} on unit test generation using Code-LLMs. They perform well on the small samples from HumanEval \cite{HumanEval} but have substandard performance on real-world open-source projects.

\paragraph*{\textbf{Guidelines for Quality Prompts}}
% Shi \etal \cite{shi2022building} used heuristic-based rules to clear noisy data of the code summarization dataset and train three models for this downstream task. Their approach improves the performance of the model. 

Shi \etal \cite{shi2022building} showed that cleaning noisy data from code summarization benchmarks can improve the performance of code summarization models.
In RQ3, we fixed the quality issues identified in the studied benchmarks' prompts. The model's performance increases a little according to the result of this research question. However, it can viewed as an updated version of the benchmark, which includes less noise, consistent, and more understandable by human prompts.  In addition, the fourth result indicates that an updated version of the existing dataset can help solve code contamination issues~\cite{sallou2023breaking}. From our result, we can suggest following: (1) Diversify the benchmark's programming languages, domains, and complexity (RQ1); (2) Use proper format and naming convention while defining the code prompts (RQ2); (3) Provide sufficient context in the prompt (RQ2); (4) Before using a sample extracted from public repositories in an evaluation set, check whether it has been used as part of training sets or not (RQ4). One way to decrease contamination likelihood is to extract code made available after the knowledge cut-off date of models; (5) Make developers aware of following a specific format while working with conversation style code generation model, as other developers can benefit from reading them (RQ5).

\paragraph*{\textbf{Data contamination and Possible Solution}}
The model can perform well if the evaluation dataset leaks in the training set. Large language models need a large amount of data to train to be generalized for different tasks \cite{brown2020language}. It can take a lot of work to deduplicate and remove test samples from the training set. In addition to that, there can be indirect leakage. For example, the HumanEval dataset was initially released by OpenAI, but it can be re-uploaded in other projects. These projects can be included in the training set, and that can be hard to capture, and may lead to data contamination. One of the possible solutions is uploading the evaluation set in a binary format \cite{jacovi2023stop}. Benchmarks can be updated regularly and benchmark the state-of-the-art models. It can also be better not to release canonical solutions to the problems, though it can create an issue to extend and verify the result. 

\paragraph*{\textbf{Implication for the Developers and Researchers}}
A code generation model's performance can affect the model's usage. Developers may prefer the model which has better performance than other ones. However, the data contamination issue indicates that the benchmark result can be rigged. Our work in RQ4 can be a way to check if the model has contamination issues. 
% In addition to that, we analyzed the developers' conversations with the ChatGPT in RQ5, and most of them are problematic (\ie around 54\% of the analyzed prompts). Hence, it may not be useful for others while going through the conversation. The prompt should be properly marked with the code and have clearly expressed intention.
Another thing is that real-world software is complex, and from RQ1, we can see that the existing benchmarks are not robust. Hence, a model can perform better on the existing evaluation dataset but may not perform well in real-world software due to hallucinations \cite{siddiq2023empirical}. 
% Researchers can work on easily finding the evaluation dataset in the training set. Though code clone detection can be handy, working on a large amount of data is hard. 
Thus, we need more robust benchmarks to evaluate the code generation model. While creating the dataset, researchers should also consider prompts that are less noisy, human-understandable, and follow standard coding practices.

\section{Threats to validity}
We manually analyzed around 3,900 prompts from developers and the benchmarks, which can introduce \textit{internal} threats to validity. However, we performed a peer review of our analyses, and Cohen's kappa score indicates a substantial agreement between the raters. Moreover,  an experienced author has resolved the disagreements. 

An \textit{external} threat to the validity is that we considered benchmarks from left to right code generation and used two versions of HumanEval datasets \cite{HumanEval,MBXP} to answer RQ3 and RQ4. However, HumanEval datasets are the \textit{most} used benchmarks for the code generation model, and the majority of the benchmarks for code generation are left-to-right \cite{zan2023NL2Code}. As we used only HumanEval datasets in RQ3 and it contained only a subset of the found issues in RQ2, the results may not be generalized for other benchmarks and with different issues.

To detect testset contamination, we detect code clones using NiCad, which introduces \textit{construct} validity threats. However, NiCad is one of the most used tools for clone detection for different languages. Another thing is that some problems in the benchmarks may have solutions that are inherently similar (\eg calculating the sum of numbers in an array) that can be solved in one way, leading to cloned solutions.

% We used three Transformer-based code generation models \cite{attention2017} as they have the best-performing result in recent times \cite{StarCoder, luo2023wizardcoder, allal2023santacoder}. 
\section{Related Work}\label{sec:RelatedWork}

% \subsection{\textbf{Empirical Study of Benchmarks}}
\subsection{{Empirical Study of Benchmarks}}

Shi \etal described a study on \textit{code summarization} benchmarks\cite{shi2022building} that characterized the data noises into 12 categories and ranked benchmarks based on the percentage of noisy data. They also developed a code-comment cleaning tool and showed that cleaning improves performance drastically up to 67.3\%. Prior works have also focused on automatically translating existing datasets to other languages~\cite{Cassano}. Cassano \etal~\cite{Cassano} translated the HumanEval and MBPP datasets into 18 different languages and compared the effectiveness of these two datasets which showed that HumanEval is more useful than MBPP. Moreover,  question-answering (QA), reasoning, and reading comprehension datasets were also evaluated based on their effectiveness \cite{zhou2023dont}.  Unlike these prior works, we studied the quality of prompts in \textbf{\textit{code generation}} benchmarks.

% A similar study on code generation benchmark datasets was needed with more emphasis on the noise detection part. As the code generation benchmark datasets come from sources very different from each other [classEval], they have different structures and conventions. A uniform structure or format is necessary for their effective interpretations. That is why we used P7, P8, P13 heuristics to judge their uniformity. A study [classEval-17] has shown that most of the methods in the datasets are part of bigger projects that heavily depend on other code contexts to be fully understood. To encounter this we evaluated the datasets based on the completeness and clarity of the prompts using P15. 

% \subsubsection*{\textbf{Empirical Study of LLMs}}
\subsection{{Empirical Study of LLMs}}
Recently, the study of LLMs has gained substantial attention owing to their good performance in various applications. 
Chang \etal~\cite{chang2023survey} presented a comprehensive survey on evaluating LLMs from three aspects: what to evaluate, where to evaluate, and how to evaluate. The authors conclude that no concrete evidence exists that one particular evaluation protocol or benchmark—albeit with distinct features and focuses—is the most beneficial and successful. They also summarized LLMs' success and failure cases in different tasks to reveal their intrinsic strengths and weaknesses.

% There has been recent empirical research on using LLMs for unit test generation \cite{siddiq2023empirical} where the authors investigated how useful three generative models (Codex, GPT-3.5-Turbo, and StarCoder) can be for unit test generation. They used three LLMs  to generate unit tests for 194 classes from 47 open-source projects in the SF110 dataset~\cite{ICSE12} and 160 classes from the HumanEval dataset \cite{HumanEval}, and the performance of the LLMs were evaluated based on metrics like branch coverage, line coverage, correctness, and quality.

Chen \etal~\cite{chen2023exploring} analyzed the effectiveness of  ChatGPT to assess the quality of the generated text. After comparing three reference-free evaluation techniques, they deduced that the Explicit Score—which uses ChatGPT to produce a numerical score indicating text quality—is the most efficient technique out of the three exploited techniques. On the other hand,  Wu \etal~\cite{wu2023empirical} evaluated the potential and constraints of different GPT-4 approaches for addressing increasingly difficult and demanding math problems.
Similarly, an assessment of ChatGPT's performance on various benchmarks has been conducted by Laskar \etal~\cite{laskar2023systematic}. They tested ChatGPT on 140 tasks and examined 255K responses produced in these datasets. Valerio \etal \cite{terragni2024future} discussed the future of AI-driven software development, specifically the requirement engineering for LLMs to understand the task. 
In our work, we focused on prompt quality for the code generation model and checked their influence on performance while fixing quality issues.

% Zhou~\etal~\cite{zhuo2023robustness} presented an empirical investigation of the adversarial robustness of a large prompt-based language model of code, Codex. They further suggest techniques to overcome the challenges of adversarial attacks on natural-language inputs by enhancing the robustness of semantic parsers. 

% \subsubsection*{\textbf{Memorization in LLMs}}
\subsection{{Memorization in LLMs}}

Carlini \etal~\cite{carlini2019secret} quantitatively measured the risk of memorization in generative sequence natural language models. 
% This methodology works as a prevention against sensitive data leaks by allowing programmers to take measures to minimize memorization.
A follow-up study showed that sensitive personal data can be easily extracted by simple attacks on a language model like GPT-2 \cite{carlini2021extracting}. Moreover, larger models are much more vulnerable to such attacks.
The model's capacity, the number of duplication of an example, and the number of tokens of context used to prompt the model demonstrate a log-linear relationship with the degree of memorization of the model \cite{carlini2023quantifying}. Though it may show unique memorization behaviors, memorization during fine-tuning was not explored much. Shorter tasks, \eg sentiment analysis and extractive QA are less likely to be memorized; on the other hand, longer tasks, \eg summarization, increase the possibility of memorization \cite{zeng2023exploring}.

%\begin{itemize}
    % \item Carlini \etal~\cite{carlini2019secret}

    % \item Carlini \etal~\cite{carlini2021extracting}

    % \item Carlini \etal~\cite{carlini2023quantifying}
    
    % \item Zeng \etal~\cite{zeng2023exploring}
    
    %\item Oren \etal~\cite{oren2023proving}
% Oren \etal~\cite{oren2023proving}  propose a \textit{exchangeability} property-based statistical method of proving test set contamination in the form of benchmark memorization in language models. They conduct a series of tests comparing the log probability of the language model on the "canonical" ordering to the log probability on a dataset containing shuffled examples and flag contamination when statistically significant differences exist between the two log probabilities. 
% Yang \etal~\cite {yang2024unveiling} investigate memorization in large pre-trained code models empirically by using the concept of code clone to determine whether output from a code model contains memorization.   
Yang \etal \cite{yang2024unveiling} studied memorization issues in open-sourced pre-trained GPT2 models. They compared the model's output to the training set, while in our study, we focused on test set contamination (\ie models memorizing the answers to existing evaluation prompts) by including not only open-source but also closed-source models (\ie GPT-3.5). 
% Moreover, a positive correlation (a correlation coefficient of 0.804 in Spearman’s rank correlation test) between the frequency of output in the training data and that in the generated outputs is shown in the study, suggesting that eliminating duplicates from the training data may be one strategy to lessen memorization.
Ippolito \etal~\cite{ippolito2022preventing} contend that definitions of verbatim memorization are overly restrictive and overlook more nuanced types of memorization. They create MEMFREE, an effective defense against all verbatim memorization. They demonstrate how this seemingly perfect filter is insufficient to protect against training data leaks.

% \paragraph{\textbf{Our Work Novelty}}
Unlike these prior works, we studied quality issues in \textbf{\textit{code generation benchmarks}} and how they affect a model's performance.
\section{Conclusion}

% Developers are sharing their conversations with the Chat-style multi-purpose model (\ie, ChatGPT) with their peers. However, our work indicates they are suffering from quality issues.

Code generation models, used to aid developers in writing code faster, are evaluated using benchmarks. In our paper, we studied these benchmarks and found that they are limited and have quality issues. Improving the quality of the benchmark can provide a better description of the prompt and may lead to a better performance of the model. 
% as shown in a survey with 500 US-based developers working for large-sized companies showed that \textbf{92\%} of them are using LLMs to generate code for work and personal use~\cite{shani2023survey}. 
In addition to that, data contamination issues can hinder the usefulness of the popular benchmark. In the future, we will explore the solution to automatically fix the prompts with quality issues, and solve data contamination issues.
% , and extend a benchmark mimicking real-world scenarios. 

\bibliographystyle{IEEEtran}
\bibliography{references}

\begin{thebibliography}{10}
\providecommand{\url}[1]{#1}
\csname url@samestyle\endcsname
\providecommand{\newblock}{\relax}
\providecommand{\bibinfo}[2]{#2}
\providecommand{\BIBentrySTDinterwordspacing}{\spaceskip=0pt\relax}
\providecommand{\BIBentryALTinterwordstretchfactor}{4}
\providecommand{\BIBentryALTinterwordspacing}{\spaceskip=\fontdimen2\font plus
\BIBentryALTinterwordstretchfactor\fontdimen3\font minus
  \fontdimen4\font\relax}
\providecommand{\BIBforeignlanguage}[2]{{%
\expandafter\ifx\csname l@#1\endcsname\relax
\typeout{** WARNING: IEEEtran.bst: No hyphenation pattern has been}%
\typeout{** loaded for the language `#1'. Using the pattern for}%
\typeout{** the default language instead.}%
\else
\language=\csname l@#1\endcsname
\fi
#2}}
\providecommand{\BIBdecl}{\relax}
\BIBdecl

\bibitem{Le_2021}
T.~H.~M. Le, H.~Chen, and M.~A. Babar, ``Deep learning for source code modeling
  and generation: Models, applications, and challenges,'' \emph{ACM Comput.
  Surv.}, vol.~53, no.~3, jun 2020.

\bibitem{perry2022users}
N.~Perry, M.~Srivastava, D.~Kumar, and D.~Boneh, ``Do users write more insecure
  code with ai assistants?'' \emph{arXiv preprint arXiv:2211.03622}, 2022.

\bibitem{shani2023survey}
\BIBentryALTinterwordspacing
I.~Shani, ``{Survey reveals AI{'}s impact on the developer experience {$\vert$}
  The GitHub Blog},'' \emph{GitHub Blog}, Jun. 2023. [Online]. Available:
  \url{https://github.blog/2023-06-13-survey-reveals-ais-impact-on-the-developer-experience/#methodology}
\BIBentrySTDinterwordspacing

\bibitem{research_copilot}
\BIBentryALTinterwordspacing
E.~Kalliamvakou, ``Research: quantifying github copilot’s impact on developer
  productivity and happiness,'' 2023, [Online; accessed 10. Nov. 2023].
  [Online]. Available:
  \url{https://github.blog/2022-09-07-research-quantifying-github-copilots-impact-on-developer-productivity-and-happiness/}
\BIBentrySTDinterwordspacing

\bibitem{albert22productivity}
A.~Ziegler, E.~Kalliamvakou, X.~A. Li, A.~Rice, D.~Rifkin, S.~Simister,
  G.~Sittampalam, and E.~Aftandilian, ``Productivity assessment of neural code
  completion,'' in \emph{Proc. of the 6th ACM SIGPLAN Int’l Symposium on
  Machine Programming}, ser. MAPS 2022.\hskip 1em plus 0.5em minus 0.4em\relax
  New York, NY, USA: ACM, 2022, p. 21–29.

\bibitem{ernst2022aide}
N.~A. Ernst and G.~Bavota, ``Ai-driven development is here: Should you worry?''
  \emph{IEEE Software}, vol.~39, no.~2, p. 106–110, Mar 2022.

\bibitem{zan2023NL2Code}
D.~Zan, B.~Chen, F.~Zhang, D.~Lu, B.~Wu, B.~Guan, Y.~Wang, and J.-G. Lou,
  ``When neural model meets {NL2Code}: A survey,'' in \emph{Proceedings of the
  61st Annual Meeting of the Association for Computational Linguistics}, 2023.

\bibitem{CoderEval}
H.~Yu, B.~Shen, D.~Ran, J.~Zhang, Q.~Zhang, Y.~Ma, G.~Liang, Y.~Li, T.~Xie, and
  Q.~Wang, ``Codereval: A benchmark of pragmatic code generation with
  generative pre-trained models,'' in \emph{International Conference on
  Software Engineering (ICSE)}, 2023.

\bibitem{carlini2019secret}
N.~Carlini, C.~Liu, {\'U}.~Erlingsson, J.~Kos, and D.~Song, ``The secret
  sharer: Evaluating and testing unintended memorization in neural networks,''
  in \emph{28th USENIX Security Symposium (USENIX Security 19)}, 2019, pp.
  267--284.

\bibitem{HumanEval}
M.~Chen, J.~Tworek, H.~Jun, Q.~Yuan, H.~P. d.~O. Pinto, J.~Kaplan, H.~Edwards,
  Y.~Burda, N.~Joseph, G.~Brockman \emph{et~al.}, ``Evaluating large language
  models trained on code,'' \emph{arXiv preprint arXiv:2107.03374}, 2021.

\bibitem{sallou2023breaking}
J.~Sallou, T.~Durieux, and A.~Panichella,
  ``\BIBforeignlanguage{English}{Breaking the silence: the threats of using
  {LLMs} in software engineering},'' in
  \emph{\BIBforeignlanguage{English}{ACM/IEEE 46th International Conference on
  Software Engineering - New Ideas and Emerging Results}}.\hskip 1em plus 0.5em
  minus 0.4em\relax ACM/IEEE, Jan. 2024.

\bibitem{carlini2023quantifying}
\BIBentryALTinterwordspacing
N.~Carlini, D.~Ippolito, M.~Jagielski, K.~Lee, F.~Tramer, and C.~Zhang,
  ``Quantifying memorization across neural language models,'' in \emph{The
  Eleventh International Conference on Learning Representations}, 2023.
  [Online]. Available: \url{https://openreview.net/forum?id=TatRHT_1cK}
\BIBentrySTDinterwordspacing

\bibitem{chatgpt}
\BIBentryALTinterwordspacing
``Chat completions,'' Accessed Mar 25, 2023, 2023. [Online]. Available:
  \url{https://platform.openai.com/docs/guides/chat}
\BIBentrySTDinterwordspacing

\bibitem{LLM}
J.~Wei, Y.~Tay, R.~Bommasani, C.~Raffel, B.~Zoph, S.~Borgeaud, D.~Yogatama,
  M.~Bosma, D.~Zhou, D.~Metzler, E.~H. Chi, T.~Hashimoto, O.~Vinyals, P.~Liang,
  J.~Dean, and W.~Fedus, ``{Emergent Abilities of Large Language Models},''
  \emph{arXiv}, Jun. 2022.

\bibitem{brown2020language}
T.~B. Brown, B.~Mann, N.~Ryder, M.~Subbiah, J.~Kaplan, P.~Dhariwal,
  A.~Neelakantan, P.~Shyam, G.~Sastry, A.~Askell, S.~Agarwal, A.~Herbert-Voss,
  G.~Krueger, T.~Henighan, R.~Child, A.~Ramesh, D.~M. Ziegler, J.~Wu,
  C.~Winter, C.~Hesse, M.~Chen, E.~Sigler, M.~Litwin, S.~Gray, B.~Chess,
  J.~Clark, C.~Berner, S.~McCandlish, A.~Radford, I.~Sutskever, and D.~Amodei,
  ``Language models are few-shot learners,'' 2020.

\bibitem{izadi2022codefill}
M.~Izadi, R.~Gismondi, and G.~Gousios, ``Codefill: Multi-token code completion
  by jointly learning from structure and naming sequences,'' in \emph{44th
  {Intl.} {Conf.} on {Software} {Engineering} ({ICSE})}, 2022.

\bibitem{kim2021code}
S.~Kim, J.~Zhao, Y.~Tian, and S.~Chandra, ``Code prediction by feeding trees to
  transformers,'' in \emph{2021 IEEE/ACM 43rd Intl. Conf. on Software
  Engineering (ICSE)}.\hskip 1em plus 0.5em minus 0.4em\relax IEEE, 2021, pp.
  150--162.

\bibitem{svyatkovskiy2021fast}
A.~Svyatkovskiy, S.~Lee, A.~Hadjitofi, M.~Riechert, J.~V. Franco, and
  M.~Allamanis, ``Fast and memory-efficient neural code completion,'' in
  \emph{2021 IEEE/ACM 18th Intl. Conf. on Mining Software Repositories
  (MSR)}.\hskip 1em plus 0.5em minus 0.4em\relax IEEE, 2021, pp. 329--340.

\bibitem{codebert}
Z.~Feng, D.~Guo, D.~Tang, N.~Duan, X.~Feng, M.~Gong, L.~Shou, B.~Qin, T.~Liu,
  D.~Jiang, and M.~Zhou, ``Codebert: A pre-trained model for programming and
  natural languages,'' \emph{arXiv preprint arXiv:2002.08155}, 2020.

\bibitem{gao2022m2ts}
Y.~Gao and C.~Lyu, ``M2ts: Multi-scale multi-modal approach based on
  transformer for source code summarization,'' in \emph{Proc. of the 30th
  IEEE/ACM Intl. Conf. on Program Comprehension}, ser. ICPC '22.\hskip 1em plus
  0.5em minus 0.4em\relax New York, NY, USA: Association for Computing
  Machinery, 2022, p. 24–35.

\bibitem{codet5}
Y.~Wang, W.~Wang, S.~Joty, and S.~C. Hoi, ``{C}ode{T}5: Identifier-aware
  unified pre-trained encoder-decoder models for code understanding and
  generation,'' in \emph{Proc. of the 2021 Conf. on Empirical Methods in
  Natural Language Processing}.\hskip 1em plus 0.5em minus 0.4em\relax Online
  and Punta Cana, Dominican Republic: Association for Computational
  Linguistics, Nov. 2021, pp. 8696--8708.

\bibitem{MBXP}
B.~Athiwaratkun, S.~K. Gouda, Z.~Wang, X.~Li, Y.~Tian, M.~Tan, W.~U. Ahmad,
  S.~Wang, Q.~Sun, M.~Shang, S.~K. Gonugondla, H.~Ding, V.~Kumar, N.~Fulton,
  A.~Farahani, S.~Jain, R.~Giaquinto, H.~Qian, M.~K. Ramanathan, R.~Nallapati,
  B.~Ray, P.~Bhatia, S.~Sengupta, D.~Roth, and B.~Xiang, ``Multi-lingual
  evaluation of code generation models,'' in \emph{The Eleventh International
  Conference on Learning Representations (ICLR)}, 2023.

\bibitem{ren2020codebleu}
S.~Ren, D.~Guo, S.~Lu, L.~Zhou, S.~Liu, D.~Tang, N.~Sundaresan, M.~Zhou,
  A.~Blanco, and S.~Ma, ``Codebleu: a method for automatic evaluation of code
  synthesis,'' \emph{arXiv preprint arXiv:2009.10297}, 2020.

\bibitem{siddiq2023generate}
M.~L. Siddiq, J.~C.~S. Santos, S.~Devareddy, and A.~Muller, ``Generate and
  pray: Using sallms to evaluate the security of llm generated code,'' 2024.

\bibitem{carlini2021extracting}
N.~Carlini, F.~Tramer, E.~Wallace, M.~Jagielski, A.~Herbert-Voss, K.~Lee,
  A.~Roberts, T.~Brown, D.~Song, U.~Erlingsson \emph{et~al.}, ``Extracting
  training data from large language models,'' in \emph{30th USENIX Security
  Symposium (USENIX Security 21)}, 2021, pp. 2633--2650.

\bibitem{zhou2023quantifying}
Z.~Zhou, J.~Xiang, C.~Chen, and S.~Su, ``Quantifying and analyzing entity-level
  memorization in large language models,'' \emph{arXiv preprint
  arXiv:2308.15727}, 2023.

\bibitem{yang2024unveiling}
Z.~Yang, Z.~Zhao, C.~Wang, J.~Shi, D.~Kim, D.~Han, and D.~Lo, ``Unveiling
  memorization in code models,'' in \emph{2024 IEEE/ACM 46th International
  Conference on Software Engineering (ICSE)}.\hskip 1em plus 0.5em minus
  0.4em\relax IEEE Computer Society, 2024, pp. 856--856.

\bibitem{ROY2009470}
C.~K. Roy, J.~R. Cordy, and R.~Koschke, ``Comparison and evaluation of code
  clone detection techniques and tools: A qualitative approach,'' \emph{Science
  of Computer Programming}, vol.~74, no.~7, pp. 470--495, 2009.

\bibitem{stol2016grounded}
K.-J. Stol, P.~Ralph, and B.~Fitzgerald, ``Grounded theory in software
  engineering research: a critical review and guidelines,'' in
  \emph{Proceedings of the 38th International conference on software
  engineering}, 2016, pp. 120--131.

\bibitem{devgpt}
T.~Xiao, C.~Treude, H.~Hata, and K.~Matsumoto, ``Devgpt: Studying
  developer-chatgpt conversations,'' in \emph{Proceedings of the International
  Conference on Mining Software Repositories (MSR 2024)}, 2024.

\bibitem{2023arXiv230200288Y}
H.~{Yu}, B.~{Shen}, D.~{Ran}, J.~{Zhang}, Q.~{Zhang}, Y.~{Ma}, G.~{Liang},
  Y.~{Li}, Q.~{Wang}, and T.~{Xie}, ``{CoderEval: A Benchmark of Pragmatic Code
  Generation with Generative Pre-trained Models},'' \emph{arXiv e-prints}, p.
  arXiv:2302.00288, Feb. 2023.

\bibitem{wang2023executionbased}
Z.~Wang, S.~Zhou, D.~Fried, and G.~Neubig, ``Execution-based evaluation for
  open-domain code generation,'' 2023.

\bibitem{MBPP}
J.~Austin, A.~Odena, M.~Nye, M.~Bosma, H.~Michalewski, D.~Dohan, E.~Jiang,
  C.~Cai, M.~Terry, Q.~Le, and C.~Sutton, ``Program synthesis with large
  language models,'' \emph{arXiv preprint arXiv:2108.07732}, 2021.

\bibitem{zan2022language}
D.~Zan, B.~Chen, Z.~Lin, B.~Guan, Y.~Wang, and J.-G. Lou, ``When language model
  meets private library,'' 2022.

\bibitem{zan2022cert}
D.~Zan, B.~Chen, D.~Yang, Z.~Lin, M.~Kim, B.~Guan, Y.~Wang, W.~Chen, and J.-G.
  Lou, ``Cert: Continual pre-training on sketches for library-oriented code
  generation,'' 2022.

\bibitem{jain2021jigsaw}
N.~Jain, S.~Vaidyanath, A.~Iyer, N.~Natarajan, S.~Parthasarathy, S.~Rajamani,
  and R.~Sharma, ``Jigsaw: Large language models meet program synthesis,''
  2021.

\bibitem{MathQA}
\BIBentryALTinterwordspacing
A.~Amini, S.~Gabriel, S.~Lin, R.~Koncel-Kedziorski, Y.~Choi, and H.~Hajishirzi,
  ``{M}ath{QA}: Towards interpretable math word problem solving with
  operation-based formalisms,'' in \emph{Proceedings of the 2019 Conference of
  the North {A}merican Chapter of the Association for Computational
  Linguistics: Human Language Technologies, Volume 1}.\hskip 1em plus 0.5em
  minus 0.4em\relax Minneapolis, Minnesota: Association for Computational
  Linguistics, Jun. 2019, pp. 2357--2367. [Online]. Available:
  \url{https://aclanthology.org/N19-1245}
\BIBentrySTDinterwordspacing

\bibitem{so2023survey}
\BIBentryALTinterwordspacing
S.~Overflow, ``Stack overflow devlopers survey,'' 2023. [Online]. Available:
  \url{https://survey.stackoverflow.co/2023/#most-popular-technologies-language-prof}
\BIBentrySTDinterwordspacing

\bibitem{mchugh2012interrater}
M.~L. McHugh, ``Interrater reliability: the kappa statistic,'' \emph{Biochemia
  medica}, vol.~22, no.~3, pp. 276--282, 2012.

\bibitem{paperswithcode}
\BIBentryALTinterwordspacing
``{Code Generation on HumanEval},'' Nov. 2023, [Online; accessed 14. Nov.
  2023]. [Online]. Available:
  \url{https://paperswithcode.com/sota/code-generation-on-humaneval}
\BIBentrySTDinterwordspacing

\bibitem{Nijkamp2022ACP}
E.~Nijkamp, B.~Pang, H.~Hayashi, L.~Tu, H.~Wang, Y.~Zhou, S.~Savarese, and
  C.~Xiong, ``A conversational paradigm for program synthesis,'' \emph{arXiv
  preprint}, 2022.

\bibitem{ThePileDataset}
L.~Gao, S.~Biderman, S.~Black, L.~Golding, T.~Hoppe, C.~Foster, J.~Phang,
  H.~He, A.~Thite, N.~Nabeshima, S.~Presser, and C.~Leahy, ``The pile: An 800gb
  dataset of diverse text for language modeling,'' 2020.

\bibitem{BigQueryDataset}
\BIBentryALTinterwordspacing
G.~Inc, ``Bigquery public datasets,'' 2022. [Online]. Available:
  \url{https://cloud.google.com/bigquery/public-data}
\BIBentrySTDinterwordspacing

\bibitem{nijkamp2023codegen2}
E.~Nijkamp, H.~Hayashi, C.~Xiong, S.~Savarese, and Y.~Zhou, ``Codegen2: Lessons
  for training llms on programming and natural languages,'' \emph{ICLR}, 2023.

\bibitem{Kocetkov2022TheStack}
D.~Kocetkov, R.~Li, L.~Ben~Allal, J.~Li, C.~Mou, C.~Muñoz~Ferrandis,
  Y.~Jernite, M.~Mitchell, S.~Hughes, T.~Wolf, D.~Bahdanau, L.~von Werra, and
  H.~de~Vries, ``The stack: 3 tb of permissively licensed source code,''
  \emph{Preprint}, 2022.

\bibitem{allal2023santacoder}
L.~B. Allal, R.~Li, D.~Kocetkov, C.~Mou, C.~Akiki, C.~M. Ferrandis,
  N.~Muennighoff, M.~Mishra, A.~Gu, M.~Dey \emph{et~al.}, ``Santacoder: don't
  reach for the stars!'' \emph{arXiv preprint arXiv:2301.03988}, 2023.

\bibitem{StarCoder}
R.~Li, L.~B. Allal, Y.~Zi, N.~Muennighoff, D.~Kocetkov, C.~Mou, M.~Marone,
  C.~Akiki, J.~Li, J.~Chim \emph{et~al.}, ``Starcoder: may the source be with
  you!'' \emph{arXiv preprint arXiv:2305.06161}, 2023.

\bibitem{luo2023wizardcoder}
Z.~Luo, C.~Xu, P.~Zhao, Q.~Sun, X.~Geng, W.~Hu, C.~Tao, J.~Ma, Q.~Lin, and
  D.~Jiang, ``Wizardcoder: Empowering code large language models with
  evol-instruct,'' 2023.

\bibitem{attention2017}
A.~Vaswani, N.~Shazeer, N.~Parmar, J.~Uszkoreit, L.~Jones, A.~N. Gomez, L.~u.
  Kaiser, and I.~Polosukhin, ``Attention is all you need,'' in \emph{Advances
  in Neural Information Processing Systems}, I.~Guyon, U.~V. Luxburg,
  S.~Bengio, H.~Wallach, R.~Fergus, S.~Vishwanathan, and R.~Garnett, Eds.,
  vol.~30.\hskip 1em plus 0.5em minus 0.4em\relax Curran Associates, Inc.,
  2017.

\bibitem{copilot}
\BIBentryALTinterwordspacing
G.~Inc., ``Github copilot : Your ai pair programmer,'' 2022, [Online; accessed
  10. Oct. 2022]. [Online]. Available: \url{https://copilot.github.com}
\BIBentrySTDinterwordspacing

\bibitem{kulal2019spoc}
S.~Kulal, P.~Pasupat, K.~Chandra, M.~Lee, O.~Padon, A.~Aiken, and P.~S. Liang,
  ``Spoc: Search-based pseudocode to code,'' in \emph{Advances in Neural
  Information Processing Systems}, H.~Wallach, H.~Larochelle, A.~Beygelzimer,
  F.~d\textquotesingle Alch\'{e}-Buc, E.~Fox, and R.~Garnett, Eds.,
  vol.~32.\hskip 1em plus 0.5em minus 0.4em\relax Curran Associates, Inc.,
  2019.

\bibitem{nicad}
J.~R. Cordy and C.~K. Roy, ``The nicad clone detector,'' in \emph{2011 IEEE
  19th International Conference on Program Comprehension}, 2011, pp. 219--220.

\bibitem{potdar2014satd}
A.~Potdar and E.~Shihab, ``An exploratory study on self-admitted technical
  debt,'' in \emph{2014 IEEE International Conference on Software Maintenance
  and Evolution}, 2014, pp. 91--100.

\bibitem{meidani2022towards}
S.~M. Meidani, ``Towards an enhanced dependency graph,'' Master's thesis,
  University of Waterloo, 2022.

\bibitem{siddiq2023empirical}
M.~L. Siddiq, J.~C.~S. Santos, R.~H. Tanvir, N.~Ulfat, F.~A. Rifat, and V.~C.
  Lopes, ``Using large language models to generate junit tests: An empirical
  study,'' in \emph{28th International Conference on Evaluation and Assessment
  in Software Engineering (EASE 2024)}, 2024.

\bibitem{shi2022building}
L.~Shi, F.~Mu, X.~Chen, S.~Wang, J.~Wang, Y.~Yang, G.~Li, X.~Xia, and Q.~Wang,
  ``Are we building on the rock? on the importance of data preprocessing for
  code summarization,'' in \emph{Proceedings of the 30th ACM Joint European
  Software Engineering Conference and Symposium on the Foundations of Software
  Engineering}, ser. ESEC/FSE 2022.\hskip 1em plus 0.5em minus 0.4em\relax New
  York, NY, USA: Association for Computing Machinery, 2022, p. 107–119.

\bibitem{jacovi2023stop}
A.~Jacovi, A.~Caciularu, O.~Goldman, and Y.~Goldberg, ``Stop uploading test
  data in plain text: Practical strategies for mitigating data contamination by
  evaluation benchmarks,'' \emph{arXiv preprint arXiv:2305.10160}, 2023.

\bibitem{Cassano}
F.~Cassano, J.~Gouwar, D.~Nguyen, S.~Nguyen, L.~Phipps-Costin, D.~Pinckney,
  M.-H. Yee, Y.~Zi, C.~J. Anderson, M.~Q. Feldman, A.~Guha, M.~Greenberg, and
  A.~Jangda, ``Multipl-e: A scalable and polyglot approach to benchmarking
  neural code generation,'' \emph{IEEE Transactions on Software Engineering},
  vol.~49, no.~7, pp. 3675--3691, 2023.

\bibitem{zhou2023dont}
K.~Zhou, Y.~Zhu, Z.~Chen, W.~Chen, W.~X. Zhao, X.~Chen, Y.~Lin, J.-R. Wen, and
  J.~Han, ``Don't make your llm an evaluation benchmark cheater,'' 2023.

\bibitem{chang2023survey}
Y.~Chang, X.~Wang, J.~Wang, Y.~Wu, L.~Yang, K.~Zhu, H.~Chen, X.~Yi, C.~Wang,
  Y.~Wang, W.~Ye, Y.~Zhang, Y.~Chang, P.~S. Yu, Q.~Yang, and X.~Xie, ``A survey
  on evaluation of large language models,'' 2023.

\bibitem{chen2023exploring}
Y.~Chen, R.~Wang, H.~Jiang, S.~Shi, and R.~Xu, ``Exploring the use of large
  language models for reference-free text quality evaluation: An empirical
  study,'' 2023.

\bibitem{wu2023empirical}
Y.~Wu, F.~Jia, S.~Zhang, H.~Li, E.~Zhu, Y.~Wang, Y.~T. Lee, R.~Peng, Q.~Wu, and
  C.~Wang, ``An empirical study on challenging math problem solving with
  gpt-4,'' 2023.

\bibitem{laskar2023systematic}
M.~T.~R. Laskar, M.~S. Bari, M.~Rahman, M.~A.~H. Bhuiyan, S.~Joty, and J.~X.
  Huang, ``A systematic study and comprehensive evaluation of chatgpt on
  benchmark datasets,'' 2023.

\bibitem{terragni2024future}
\BIBentryALTinterwordspacing
V.~Terragni, P.~Roop, and K.~Blincoe, ``{The Future of Software Engineering in
  an AI-Driven World},'' in \emph{Workshop 2030 Software Engineering co-located
  with FSE 2024}, 2024. [Online]. Available:
  \url{https://arxiv.org/abs/2406.07737}
\BIBentrySTDinterwordspacing

\bibitem{zeng2023exploring}
S.~Zeng, Y.~Li, J.~Ren, Y.~Liu, H.~Xu, P.~He, Y.~Xing, S.~Wang, J.~Tang, and
  D.~Yin, ``Exploring memorization in fine-tuned language models,'' \emph{arXiv
  preprint arXiv:2310.06714}, 2023.

\bibitem{ippolito2022preventing}
D.~Ippolito, F.~Tram{\`e}r, M.~Nasr, C.~Zhang, M.~Jagielski, K.~Lee, C.~A.
  Choquette-Choo, and N.~Carlini, ``Preventing verbatim memorization in
  language models gives a false sense of privacy,'' \emph{arXiv preprint
  arXiv:2210.17546}, 2022.

\end{thebibliography}

\end{document}